%% file: main.tex
\documentclass[superscriptaddress,
aps,
]{revtex4-2}
\usepackage{graphicx}
\usepackage{dcolumn}
\usepackage{bm}
\usepackage[utf8]{inputenc}
\usepackage{amssymb}
\usepackage{amsmath}
\usepackage{braket}
\usepackage{hyperref}
\usepackage{mathrsfs}
\usepackage{mathtools}
\usepackage[title]{appendix}
\usepackage{glossaries}
\usepackage{dsfont}
\usepackage{qcircuit}
\usepackage[caption=false]{subfig}
\usepackage{textcomp}
\usepackage{orcidlink}
\usepackage{mhchem}

\DeclareMathOperator{\e}{e}
\patchcmd{\appendices}{\quad}{:\quad}{}{}

\newcommand{\lspan}{\mathrm{span}}

\newacronym{qsp}{QSP}{quantum signal processing}
\newacronym{qpe}{QPE}{quantum phase estimation}
\newacronym{qpu}{QPU}{quantum processing unit}
\newacronym{qksd}{QKSD}{quantum Krylov subspace diagonalization}
\newacronym{vqe}{VQE}{variational quantum eigensolver}
\newacronym{rdm}{RDM}{reduced density matrice} 
\newacronym{lcu}{LCU}{linear combination of unitaries}

\begin{document}

\title{Molecular Properties from Quantum Krylov Subspace Diagonalization}
\date{\today}

\author{Oumarou Oumarou~\orcidlink{0009-0007-5517-5229}}
\affiliation{
Covestro Deutschland AG, Leverkusen 51373, Germany
}

\author{Pauline J. Ollitrault~\orcidlink{0000-0003-1351-7546}}
\affiliation{QC Ware Corp, Palo Alto, USA and Paris, France}

\author{Cristian L. Cortes~\orcidlink{0000-0002-1163-2981}}
\affiliation{QC Ware Corp, Palo Alto, USA and Paris, France}

\author{Maximilian Scheurer~\orcidlink{0000-0003-0592-3464}}
\affiliation{
Covestro Deutschland AG, Leverkusen 51373, Germany
}

\author{Robert M. Parrish~\orcidlink{0000-0002-2406-4741}}
\affiliation{QC Ware Corp, Palo Alto, USA and Paris, France}

\author{Christian Gogolin~\orcidlink{0000-0003-0290-4698}}
\affiliation{
Covestro Deutschland AG, Leverkusen 51373, Germany
}

\begin{abstract}
    Quantum Krylov subspace diagonalization is a prominent candidate for early fault tolerant quantum simulation of many-body and molecular systems, but so far the focus has been mainly on computing ground-state energies.
    We go beyond this by deriving analytical first-order derivatives for quantum Krylov methods and show how to obtain relaxed one and two particle reduced density matrices of the Krylov eigenstates.
    The direct approach to measuring these matrices requires a number of distinct measurement that scales quadratically with the Krylov dimension $D$. Here, we show how to reduce this scaling to a constant. This is done by leveraging quantum signal processing to prepare Krylov eigenstates, including exited states, in depth linear in $D$. 
    We also compare several measurement schemes for efficiently obtaining the expectation value of an operator with states prepared using quantum signal processing. 
    We validate our approach by computing the nuclear gradient of a small molecule and estimating its variance.  
\end{abstract}

\maketitle

\section{Introduction}

Predicting the energies and properties of complex, large molecules remains a challenging yet crucial goal. 
The advent of quantum computers has introduced a fundamentally new approach to solving this problem, promising feasible time scales for determining the ground-state energy of industrially significant molecules, such as P450 or FeMoco, through the \gls{qpe} algorithm~\cite{reiher2017elucidating,babbush2018encoding, berry2019qubitization,lee2021even, su2021fault, kim2022fault, delgado2022simulating, von2021quantum, goings2022reliably, rocca2024reducing}.
While these estimates are promising, they still require substantial quantum resources that are not yet available.
Over the past decade, considerable effort has been devoted to minimizing these resource requirements.
A prominent direction has been the optimization of key components of \gls{qpe}, including state initialization~\cite{lee2023evaluating,marti2024spin,ollitrault2024enhancing,morchen2024classification}, Hamiltonian simulation algorithms~\cite{berry2015hamiltonian, low2017optimal, low2019hamiltonian}, the reduction of ancilla qubits~\cite{griffiths1996semiclassical,higgins2007entanglement,lin2022heisenberg,ding2023even}, or optimizing circuit depth and the energy estimator~\cite{dutkiewicz2024errormitigationcircuitdivision}.
It is worth noting that the literature on these topics is extensive; here, we cite only a selection of representative works for each of the mentioned area.

Alternatively, efforts to minimize resource requirements have driven the development of entirely new algorithms.
These alternatives often exchange \gls{qpe}'s substantial circuit depth for additional incoherent steps and measurement processes, making them more suitable for near-term quantum devices~\cite{bharti2022noisy,tong2022designing}.
One particularly promising approach is \gls{qksd}, which maps the Hamiltonian onto a low-dimensional Krylov subspace~\cite{parrish2019quantum, seki2021quantum, motta2020determining}. This method reduces the computational demands while maintaining the accuracy required for practical applications.
The quantum computer is used to obtain the matrix elements of the Hamiltonian within this Krylov basis, and a low-dimensional generalized eigenvalue problem is then solved on a classical computer to find the best approximation of the lowest-lying eigenstates and their corresponding energies.

This approach is advantageous for several reasons.
First, it requires shallower quantum circuits than \gls{qpe} while avoiding the challenging parameter optimization problem inherent to other variational methods, such as the \gls{vqe}~\cite{peruzzo2014variational, mcclean2018barren}.
The accuracy of the ground-state approximation is also known to converge rapidly with the subspace dimension ~\cite{saad1980rates,epperly2022theory}.
Moreover, the Hamiltonian and overlap matrices within the Krylov subspace exhibit a Hankel structure, meaning that they are defined by a number of elements that grows linearly with subspace dimension.
These last two features effectively reduce the number of calls to the \gls{qpu}.
Leveraging these benefits, this algorithm has been successfully implemented on a quantum computer with up to $57$ qubits, one of the largest quantum computing experiments to solve a quantum many-body problem to date~\cite{yoshioka2024diagonalization}.

The most recent theoretical studies of \gls{qksd} have largely focused on understanding the impact of noise on the stability and accuracy of the quantum Krylov methods.
Ref.~\cite{Kirby_2024} addresses error bounds for real-time Krylov algorithms.
A linear scaling of ground-state energy error with input noise is demonstrated and methods to manage variational violations are proposed.
Complementing this, Ref.~\cite{Lee_2024} investigates the impact of finite sampling errors in \gls{qksd}, proposing strategies such as optimal thresholding to mitigate ill-conditioning and improve algorithm resilience.
Numerical simulations of the one-dimensional Hubbard model underscores the method's practical accuracy and error predictability.

One point that remains absent from the literature is the calculation of energy derivatives in \gls{qksd}, which is crucial for extracting molecular properties beyond just energy values.
An important example, which will be the focus of this work, is the first order derivative of the energy with respect to the nuclear coordinates.
It enables analyses of the potential energy surface such as geometry optimization or reaction pathway simulation~\cite{bratovz1958calcul,gerratt1968force,pulay1969ab, kato1979energy, goddard1979gradient, pulay1979systematic, brooks1980analytic, krishnan1980derivative, dupuis1981energy, nakatsuji1981force, schaefer1986robert}.
While computing such derivatives using finite differences is straightforward, this approach often suffers from inefficiency and inaccuracy due to the need for multiple evaluations and susceptibility to numerical instability.
This makes analytical approaches essential for practical and accurate quantum chemistry simulations~\cite{kassal2009quantum, parrish2019hybrid, hohenstein2022efficient}.
Analytical derivatives have been studied in the context of \gls{qpe}~\cite{kassal2009quantum, o2019calculating, o2022efficient}, \gls{vqe}~\cite{o2019calculating, mitarai2020theory, parrish2021analytical, hohenstein2022efficient, o2022efficient} and in extensions of \gls{vqe} to excited states~\cite{parrish2019hybrid, parrish2021analytical}.

In this work, we derive a theoretical framework for obtaining analytical nuclear energy gradients with \gls{qksd} via the one and two-particle \glspl{rdm}.
We consider a Krylov basis composed of monomial functions of the Hamiltonian, i.e., spanned by applying powers of the Hamiltonian to a fixed reference state.
We show that while the number of calls to the \gls{qpu} scales linearly with the subspace dimension for energy calculations, obtaining a single \gls{rdm} element naively requires a quadratic number of \gls{qpu} executions in the Krylov space dimension.
We propose an alternative approach that reduces the number of calls to the \gls{qpu} while keeping the same circuit depth as required for computing energies and discuss its practical implementation. 

The rest of this paper is structured as follows:
Section~\ref{sec:background} begins by deriving the expressions for \gls{qksd} energies and gradients, highlighting the quadratic increase in resources required for gradient estimation.
We then discuss a method to circumvent this scaling by directly preparing the \gls{qksd} ground-state, detailing the practical implementation using \gls{qsp} in Section~\ref{sec:krylov_state_prep}.
In Section~\ref{sec:rdm_evaluation}, we explore two approaches for measuring the \glspl{rdm} on the prepared \gls{qksd} ground-state.
Finally, in Section~\ref{sec:results}, we compare the efficiency of these measurement methods through numerical simulations.

\section{Quantum Krylov Energy and Gradients}
\label{sec:background}
\subsection{Energies in the Krylov subspace}
Let $\hat{H}$ be a given Hamiltonian and $\ket{\psi_0}$ be a fixed reference state such that $\ket{\psi_0}$ is not orthogonal to the eigenstates of $\hat{H}$ that are of interest, e.g., the ground-state.
We define the Krylov subspace $\mathcal{K}$ of dimension $D$ as the linear span $\mathcal{K}$ of the Krylov states ${\ket{\psi_i}, i=0,\dots,D-1}$.
The Krylov subspace is then defined as: 
\begin{align}
    \label{eq:monomials}
    \ket{\psi_k}& \coloneqq \hat{H}^k\ket{\psi_0}\\
    \mathcal{K}& \coloneqq \lspan(\{\ket{\psi_k}\}_{k=0}^{D-1}) .
\end{align}
The Krylov space spanned by monomials is equivalent to the Krylov space spanned by any family of polynomials of the Hamiltonian of increasing degree that contains all powers up to the maximum power, such as Chebychev polynomials \cite{kirby2023exact}, which we will use later. 
This choice is motivated by the fact that firstly, these polynomial functions can be prepared with no approximation error at finite depth using \gls{qsp}.
Secondly, differentiating the Krylov energy in any polynomial basis reduces at the end to differentiating monomials. 

There are alternative ways in which the Krylov space can be constructed.
Ref.~\cite{parrish2019quantum} for instance uses time-evolution under the Hamiltonian of the reference state $\ket{\psi_0}$ for different times $k\Delta t$ such that $\ket{\psi_k} \coloneqq e^{-ikH\Delta t}\ket{\psi_0}$.
In a similar fashion, in Ref.~\cite{motta2020determining} the Krylov states are generated using imaginary time evolution.
Ideally, one would want the matrix $\bra{\psi_i} \hat{H} \ket{\psi_j}, 0\leq i,j \leq D-1$ to be a Toeplitz matrix, but if approximate, i.e., trotterized, (imaginary) time evolution is used, the Toeplitz property is not necessarily preserved. 
Other functions of the Hamiltonian that are beyond time evolution (real or imaginary) have also been applied to generate the Krylov subspace~\cite{Zhang_2024}.
The choice of one or the other is guided by their inherent properties such as number of samples needed for a target accuracy, the complexity of preparing and determining expectation values and overlaps between the Krylov states.


Let $\hat H = \sum_j \lambda_j \ket{\lambda_j}\bra{\lambda_j}$ be the spectral decomposition of $\hat H$ with $\lambda_j$ (possibly degenerate) and ordered in ascending order.
Given an appropriately chosen reference state $\ket{\phi_0}$, the ground-state energy, 
and in fact also subsets of excited state energies, are approximated by eigenvalues $E_m$ within the Krylov subspace $\mathcal{K}$. 
These energies can be analytically determined by solving the generalized eigenvalue problem 
\begin{equation}
    \widetilde{H}c^m=E_m\widetilde{S}c^m
    \label{eq:gen_eigen}
\end{equation} 
where 
\begin{align}
    \widetilde{H}_{ij}& \coloneqq \bra{\psi_i}\hat{H}\ket{\psi_j}=\bra{\psi_0}\hat{H}^{i+j+1}\ket{\psi_0}\\
    \widetilde{S}_{ij}& \coloneqq \langle\psi_i |\psi_j\rangle=\bra{\psi_0}\hat{H}^{i+j}\ket{\psi_0}=\widetilde{H}_{i,j-1}.
\end{align}
Since the overlap matrix 
$\widetilde{S}=U \Lambda U^\dagger= (U\sqrt{\Lambda})(U\sqrt{\Lambda})^{\dagger}$ is symmetric and positive definite, the lowest energy in the Krylov subspace is equivalently the lowest eigenvalue of the canonical orthonormalized generalized eigenvalue problem~\cite{lowdin1956quantum, lowdin1970nonorthogonality}:
\begin{equation}
    \label{eq:can_eig}
    (( U\sqrt{\Lambda}^{-1})^{\dagger} \widetilde{H}  U\sqrt{\Lambda}^{-1})y^m=E_my^m
\end{equation} 
We can thus obtain the eigenstates $c^m = U\sqrt{\Lambda}^{-1}y^m$ and eigenvalues
\begin{align}
    E_m =& \sum_{i,j}c_i^mc^m_j\widetilde{H}_{ij},
\end{align}
of the generalized eigenvalue problem by solving \eqref{eq:can_eig}.


The matrices $\widetilde{H}$ and $\widetilde{S}$ are of Hankel form, i.e., their rows are related by a shift as in the matrix
\begin{equation}
    A = \begin{pmatrix}
        a_0 & a_1 & a_2 & \cdots & a_{n-1} \\
        a_1 & a_2 &     &        & \vdots \\
        a_2 &     &     &        & a_{2n-4} \\
        \vdots &  &     & a_{2n-4} & a_{2n-3} \\
        a_{n-1}   & \cdots &  a_{2n-4} & a_{2n-3}  & a_{2n-2} \\
    \end{pmatrix}.
\end{equation}
Therefore, $\widetilde{H}$ and $\widetilde{S}$ are completely defined by $2D-1$ and $2D-2$ entries, respectively (since $\widetilde{S}_{00} = 1$ by definition). 
This implies that only a linear (in $D$) number of distinct quantities need to be measured on the quantum device to obtain energies via the generalized eigenvalue problem. 
This is a great strength of the quantum Krylov approach.
However, as we shall see, this linear scaling in $D$ is lost when trying to obtain energy derivatives in a straightforward way.

\subsection{Derivatives of the Krylov energy}
\label{subsec:deri_krylov_energy}
From Eq.~\eqref{eq:gen_eigen} the expression of the derivative of $E_0$  with respect to any parameter $\theta$ defining the Hamiltonian is (see Appendix~\ref{app:proofs} for a proof)
\begin{equation}
\label{eq:general_derivative_of_gs_energy}
    \frac{d E_0}{d \theta} =\frac{1}{\sum_{i,j=0}^{D-1}c^0_ic^0_j\widetilde{S}_{ij}} \sum_{ij} c^0_ic^0_j \left( \frac{d \widetilde{H}_{ij}}{d \theta} - E_0\frac{d \widetilde{S}_{ij}}{d \theta} \right).
\end{equation}
Explicitly writing the dependence of the Hamiltonian on the parameter $\theta$ as $\hat{H} = \hat{H}(\theta)$ and using the product rule we get
\begin{equation}
\label{eq:derivative_of_theta_dependent_hamiltonian}
    \frac{d \widetilde{H}_{ij}}{d \theta}= \sum_{k=1}^{i+j+1} \bra{\psi_0} \hat{H}(\theta)^{k-1} \frac{\partial \hat{H}(\theta)}{\partial \theta} \hat{H}(\theta)^{i+j+1-k}\ket{\psi_0} .
\end{equation}
Note that to simplify the notation, we only included the Hellmann-Feynman term and omitted the response terms stemming from the Hartree-Fock molecular orbital variations as these are well established and re-deriving them is out of the scope of this work (see Ref.~\cite{hf_grad} and Section~D of Appendix~$\MakeUppercase{\romannumeral 3}$ of Ref.~\cite{hohenstein2022efficient}).
If $\partial \hat{H}(\theta)/\partial \theta$ and $\hat{H}(\theta)$ do not commute, $\partial \widetilde{H}_{ij}/\partial \theta$ is still a Hankel matrix, but the $i,j$-th matrix element now requires evaluating the expectation value of $i+j$ distinct non-hermitian operators. 
A similar observation is made for the matrix elements $\partial \widetilde{S}_{ij}/\partial \theta$, since $\widetilde{S}_{ij}=\widetilde{H}_{i,j-1}$.
This leaves one with the unsatisfying situation of having to measure $\mathcal{O}(D^2)$ distinct expectation values to evaluate gradients with the quantum device.
Since precision is expected to increase exponentially with $D$ \cite{koch20118} (albeit with potential impact from sampling noise) one one might assume this renders the scaling issue negligible. However, for practically relevant $D$, typically around 10 to 20, the difference between linear scaling and quadratic scaling  becomes significantly pronounced and cannot be overlooked.

Here we are particularly interested in obtaining the nuclear gradients, i.e., the derivatives of the molecular electronic ground-state energy with respect to the position of the nuclei or, in other words, the forces acting on the atoms.
In this case, the Hamiltonian is of the form 
\begin{equation}
    \hat{\mathcal{H}}_\mathrm{mol} \coloneqq E_{\textrm{nuc}} + \sum_{pq}^N k_{pq} \hat{E}_{pq} +
    \frac{1}{2} \sum_{pqrs}^N g_{pqrs} \hat{E}_{pq} \hat{E}_{rs},
    \label{eq:est_hamiltonian}
\end{equation}
where $N$ is the number of spin orbitals, and the one-particle excitation operators are defined as $\hat{E}_{pq}\coloneqq 
\hat{a}^{\dag}_p \hat{a}_q$, with $\hat{a}^{\dag}$ and $\hat{a}$ as the fermionic creation and annihilation operators, respectively. 
The constant term, $E_{\textrm{nuc}}$, corresponds to the nuclear repulsion energy, $g_{pqrs}$ denotes the two-electron integrals,
and $k_{pq}$ is the effective one-body operator \cite{hohenstein2022efficient}.
The nuclear gradient can then be expressed using the chain rule,
\begin{align}
    \frac{d E_0}{d x}=\sum_{pq}\frac{d E_0}{d k_{pq}}\frac{d k_{pq}}{d x}+\sum_{pqrs}\frac{d E_0}{d g_{pqrs}}\frac{d g_{pqrs}}{d x}.
    \label{eq:nucl_grad_chain_rule}
\end{align}
The expression for $dE_0/dk_{pq}$ can be found from Eq.~\eqref{eq:general_derivative_of_gs_energy}, and Eq.~\eqref{eq:derivative_of_theta_dependent_hamiltonian} can also be rewritten as
%
\begin{align} \label{eq:derivative_of_k_dependent_hamiltonian}
    \frac{d \widetilde{H}_{ij}}{d k_{pq}}= \sum_{k=1}^{i+j+1} \bra{\psi_0} \hat{H}^{k-1} \hat{E}_{pq} \hat{H}^{i+j+1-k}\ket{\psi_0}.
\end{align}
Note that similar expressions can be obtained for the derivatives with respect to the two-electron integrals. 
To avoid performing $\mathcal{O}(D^2)$ distinct overlap measurements $\bra{\psi_0}H^k\hat{E}_{pq}H^j\ket{\psi_0}$, one can instead prepare the quantum Krylov ground-state $\ket{\Psi_0}=\sum_{i=0}^{D-1}c_i^0\hat{H}^i\ket{\psi_0}$ on the \gls{qpu} and use the Hellmann-Feynman theorem to replace Eq.~\eqref{eq:general_derivative_of_gs_energy} by
\begin{equation}\label{eq:derivative_1rdm}
    \gamma_{pq} := \frac{d E_0}{d k_{pq}} = \braket{\Psi_0 | \hat{E}_{pq} | \Psi_0}
\end{equation}
and 
\begin{equation}\label{eq:derivative_2rdm}
    \Gamma_{pqrs} := \frac{d E_0}{d g_{pqrs}} = \braket{\Psi_0 | \hat{E}_{pq}\hat{E}_{rs} | \Psi_0}.
\end{equation}
Indeed, preparing $\ket{\Psi_0}$ with \gls{qsp} has an inherent success probability $p_{\text{success}} < 1$ that depends on several factors such as the dimension of the Krylov space $D$. 
Another important consideration is the nature of the polynomial itself. Polynomials that are ill-conditioned or degenerate can be challenging particularly using methods that require roots solving of high degree polynomials \cite{dong2021efficient}. 
We proceed with this approach for the calculation for the 1- and 2-body \glspl{rdm} mainly because, as previously mentioned, the number of distinct measurements needed reduces from $O(D^2)$ to $O(1)$.
Regarding the variance properties we benchmark the \gls{qsp} based approach and one would need similar benchmarks for the verbose overlap approach to comment on how the two approaches compare.
%
%
In the following section we detail how to prepare $\ket{\Psi_0}$ in practice.
In Section~\ref{sec:rdm_evaluation} we then show how to efficiently evaluate $\gamma_{pq}$ and $\Gamma_{pqrs}$.


\section{Quantum Krylov eigenstate preparation}
\label{sec:krylov_state_prep}
As stated above we can prepare $\ket{\Psi_0}$, resulting from the quantum Krylov algorithm, to bypass the $\mathcal{O}(D^2)$ term measurements needed to calculate the nuclear gradients.
This can be achieved with \gls{qsp}~\cite{low2019hamiltonian, o2022efficient, Motlagh_2024}.
In this section, we first present the foundational oracles used to block-encode a Hamiltonian in its \gls{lcu} form.
Then we present the \gls{qsp} circuit and detail how to generate the appropriate gate parameters to block-encode a given polynomial. 


\subsection{Block encoding and Qubitization}
The electronic structure Hamiltonian expressed in a molecular orbital basis, after mapping to a $N$-qubit space (system register) via the Jordan-Wigner mapping, can be written as a linear combination of products $\hat{P}_k$ of Pauli operators
\begin{equation}
    \hat{\mathcal{H}} = \sum_k \alpha_k \hat{P}_k,
\end{equation}
where the coefficients $\alpha_k$ are real.
Without loss of generality, we can consider that all $\alpha_k \geq 0$ by absorbing phase factors into the $\hat{P}_k$ and we use the subscripts $a$ and $s$ to mark the ancillary and system register qubits. 
The operator $\hat{G}$ acts on an ancilla register as
\begin{equation}
    \hat{G} \ket{0}_a := \ket{G}_a = \sum_k \sqrt{\frac{\alpha_k}{\lambda_{\mathrm{LCU}}}} \ket{k}_a
\end{equation}
where $\lambda_{\mathrm{LCU}} \coloneqq \sum_k \alpha_k$.

The operator $\hat{U}$ is then defined on the full set of qubits (system and ancilla registers) as
\begin{equation}
    \hat{U}\ket{k}_a\ket{\psi_0}_s = \ket{k}_a \hat{P}_k \ket{\psi_0}_s.
\end{equation}
Combining these operators, we obtain the block encoding of $\hat{H}\coloneqq \hat{\mathcal{H}}/\lambda_{\mathrm{LCU}}$ in the subspace where all the ancilla qubits are in the $\ket{0}$ state,
\begin{equation}
    \bra{0}_a \hat{U}_H \ket{0}_a\ket{\psi_0}_s = \bra{G}_a \hat{U} \ket{G}_a\ket{\psi_0}_s = \hat{H} \ket{\psi_0}_s.
\end{equation}
where $\hat{U}_H \coloneqq \hat{G}^{\dagger}\hat{U}\hat{G}$.
Given the spectral decomposition 
\begin{equation}
    \hat{H}= \sum_j \lambda_j \ket{\lambda_j}_s\bra{\lambda_j}_s
\end{equation}
with $\lambda_j \in [-1,1]$, we have that for each eigenstate $\ket{\lambda_j}$
\begin{equation}
    \hat{U} \ket{G}_a\ket{\lambda_j}_s = \lambda_j \ket{G}_a\ket{\lambda_j}_s + \sqrt{1-\lambda_j^2} \ket{G_j^{\bot}}
\end{equation}

Since $\hat{U}$ is the product of controlled-$\hat{P}_k$ operators, it is self inverse, i.e $\hat{U}^2 = \hat{\mathds{1}}$. 
Therefore, by defining the reflection around $\ket{G}$
\begin{equation}
    \label{eq:reflection}
    \hat{R}_{G} \coloneqq (2 \ket{G}_a\bra{G}_a - \hat{\mathds{1}}_a)\otimes \hat{\mathds{1}}_s,
\end{equation}
we know from Corollary~9 of Ref.~\cite{low2019hamiltonian}, that the iterate $\hat{W}=\hat{R}_{G}\hat{U}$ leaves the two-dimensional subspaces $\mathcal{B}_j$ spanned by the state $\ket{G}_a\ket{\lambda_j}_s$ and $\hat{W}\ket{G}_a\ket{\lambda_j}_s$ invariant, i.e., $\text{span}(\ket{G}_a\ket{\lambda_j}_s, \hat{W}\ket{G}_a\ket{\lambda_j}_s)=\text{span}(\ket{G}_a\ket{\lambda_j}_s, \ket{G^\perp_j})$. 
In this two-dimensional basis the iterate reads
\begin{equation}
    [\hat{W}]_{\mathcal{B_j}} = \begin{pmatrix}
    \lambda_j & -\sqrt{1-\lambda_j^2}\\
    \sqrt{1-\lambda_j^2} & \lambda_j
\end{pmatrix}.
\end{equation}
We see that $[\hat{W}]_{\mathcal{B_j}} = \hat{R}_Y(2\arccos(\lambda_j)) = e^{i \arccos(\lambda_j) \hat{\sigma}_Y}$. Multiple applications of the iterate therefore result in a highly structured behavior.
This concept, called qubitization, is used in \gls{qsp} to construct the block-encoding of a wide set of polynomial functions.

\subsection{Quantum Signal Processing}
\label{sec:qsp}
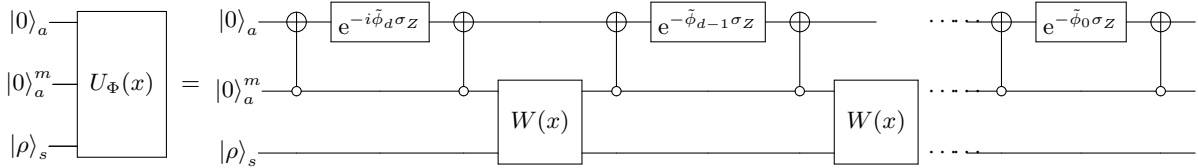
\begin{figure}
\begin{align*}
        \begin{array}{c}
        \Qcircuit @C=1em @R=1.5em {
            \push{\ket{0}_a} & \multigate{2}{U_{\Phi}(x)} \qw\\
            \push{\ket{0}^{m}_a} & \ghost{U_{\Phi}(x)} \qw\\
            \push{\ket{\rho}_s} & \ghost{U_{\Phi}(x)} \qw
            }
        \end{array} 
        =\begin{array}{c}
        \Qcircuit @C=1em @R=1.5em {
            \push{\ket{0}_a}& \targ & \gate{\e^{-i\tilde{\phi}_d \sigma_Z}} & \targ & \qw & \targ & \gate{\e^{-\tilde{\phi}_{d-1}\sigma_Z}} & \targ & \qw & \cdots &  \cdots & \targ & \gate{\e^{-\tilde{\phi}_0 \sigma_Z}} & \targ & \qw\\
            \push{\ket{0}^{m}_a}& \ctrlo{-1} & \qw & \ctrlo{-1} & \multigate{1}{W(x)} & \ctrlo{-1} & \qw & \ctrlo{-1} & \multigate{1}{W(x)}
            & \cdots & \cdots & \ctrlo{-1} & \qw & \ctrlo{-1} & \qw\\
            \push{\ket{\rho}_s}& \qw & \qw & \qw & \ghost{W(x)} & \qw & \qw & \qw & \ghost{W(x)} & \cdots &  \cdots & \qw & \qw & \qw & \qw
            }
        \end{array}
    \end{align*}
    \caption{Quantum circuit for block-encoding a polynomial $P(x)$ of fixed parity using quantum signal processing as defined in Eq.~\eqref{eq:qsp}.}
    \label{fig:circiut_1}
\end{figure}

In what follows we define $\hat{W}(x)\coloneqq e^{i\arccos(x)\hat{\sigma}_Y}$.
The \gls{qsp} theorem~\cite{low2019hamiltonian} states that given an integer $d>0$ and two complex polynomials $P(x)$ and $Q(x)$ satisfying the following conditions,
\begin{enumerate}
    \item $\deg(P)\leq d$, $\deg(Q)\leq d-1$,
    \item parity of $P=d\mod 2$, parity of $Q = d-1 \mod 2$ and
    \item $|P(x)|^2+(1-x^2)|Q(x)|^2 = 1$,
\end{enumerate} 
there exist phase factors $\Phi = (\phi_0,\dots,\phi_d) \in [-\pi,\pi)^{d+1}$ such that
\begin{align}
    \hat{U}_\Phi(x)= e^{i (\phi_0 - \frac{\pi}{4}) \hat{\sigma}_z} \Big( \prod_{k=1}^{d-1} \hat{W}(x) \e^{i\phi_k \hat{\sigma}_z} \Big) \e^{i(\phi_d + \frac{\pi}{4}) \hat{\sigma}_z}=\begin{pmatrix}
        P(x) & i\sqrt{1-x^2}Q(x)\\
        i\sqrt{1-x^2}Q^*(x) & P^*(x)
    \end{pmatrix} .
    \label{eq:qsp}
\end{align}
The corresponding circuit to $\hat{U}_\Phi(x)$ in Eq.~\eqref{eq:qsp} is laid out in Figure~\ref{fig:circiut_1}, in which, for improved readability, we have defined $\tilde{\phi}_i\coloneqq\phi_i, \; i=1,\dots,d-1$, $\tilde{\phi}_0\coloneqq\phi_0-\pi/4$ and $\tilde{\phi}_d\coloneqq\phi_d+\pi/4$.

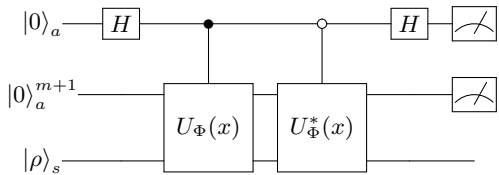
\begin{figure}
    \begin{align*}
        \begin{array}{c}\label{fig:circiut}
        \Qcircuit @C=1em @R=1.5em {
            \push{\ket{0}_a}& \gate{H} & \ctrl{1} & \ctrlo{1} & \gate{H} & \meter \\
            \push{\ket{0}^{m+1}_a}& \qw & \multigate{1}{U_\Phi(x)} & \multigate{1}{U^*_\Phi(x)}  & \qw & \meter \\
            \push{\ket{\rho}_s}& \qw  & \ghost{U_\Phi(x)} & \ghost{U^*_\Phi(x)} & \qw & \qw
            }
        \end{array}
    \end{align*}
    \caption{Quantum circuit template for block-encoding the real part of $P(x)$ presented in Eq.~\eqref{eq:qsp}. where $U_\Phi$ the circuit depicted in Figure~\ref{fig:circiut_1} for block-encoding $P$ and $P^*$ respectively. 
    \label{fig:circuit}
    }
\end{figure}

While in general \gls{qsp} applies to complex polynomial, we restrict ourselves to the real case since the electronic structure Hamiltonian and thus the Krylov space and ground-state are real.
To construct the block-encoding of a general real polynomial $f(x)$, the polynomial is first decomposed into its odd and even components: $f=1/2(f_{\text{odd}}+f_{\text{even}})$.
Note that the explicit dependence on $x$ is now omitted to simplify the notation.
The next step involves determining the complex polynomials $P_{\text{even(odd)}}$ and $Q_{\text{even(odd)}}$ that satisfy the conditions outlined above, ensuring that the real part of $P_{\text{even(odd)}}$ corresponds to $f_{\text{odd(even)}}$.
Note that the existence of such polynomials is guaranteed~\cite{gilyen2019quantum}. 
Then the phase factors, $\Phi$, must be determined either analytically or numerically. 
In the numerical approach \cite{dong2021efficient}, $\Phi$ is obtained by minimizing $||\bra{0}U_\Phi\ket{0} - P_{\text{odd(even)}}||$ over a grid of points $x_k=\cos(\frac{(2j-1)\pi}{4\tilde{d}})$ where $\tilde{d}=\lceil\frac{d+1}{2}\rceil$ and $j \in [1,\tilde{d}]$.
This method is indeed more numerically stable for high degree cases compared to the analytical approaches \cite{gilyen2019quantum} as it does not require root finding methods. 
Once the phase factors are determined, the block-encoding of $f_{\text{odd(even)}}$ is constructed by introducing an extra ancilla qubit initialized in the $\ket{+}$ state to control the unitaries $U_{\Phi}$ and $U_{\Phi}^*$ as illustrated in Figure~\ref{fig:circuit}).
This produces $1/2(P_{\text{odd(even)}}+P^*_{\text{odd(even)}})=f_{\text{odd(even)}}$. 
Finally, the block-encoding of $f$ is obtained by adding an extra qubit and by applying the same procedure to combine the block-encodings of $f_{\text{odd}}$ and $f_{\text{even}}$. 

\section{Evaluation of gradients on the QSP prepared eigenstate}
\label{sec:rdm_evaluation}
%
For the rest of the manuscript, we use Chebyshev polynomials as our Krylov basis as the Chebyshev basis is more convenient in the context of qubitization.
As previously mentioned, both the monomials of Eq.~\eqref{eq:monomials} and the Chebyshev polynomials span the same subspace and therefore the Krylov method yields the same ground-state in both cases. 
The straightforward differentiation of the Krylov energy also produces the same increase of complexity to a $O(D^2)$ scaling.
By solving the generalized eigenvalue problem in the Krylov subspace one obtains the eigenenergies and the coefficients $c^m_i$ of the eigenstates. Without loss of generality, we proceed next by fixing $m=0$ i.e. to the ground-state and energy approximation case.
The quantum state corresponding to the Krylov ground-state energy is then given by $\ket{\Psi_0}=\sum_{i=0}^{D-1} c^0_i T_i(\hat{H})\ket{\psi_0}$.
The function $\sum_{i=0}^{D-1}c_i^0T_i(H)$ needs to be appropriately normalized before \gls{qsp} can be applied~\cite{dong2021efficient}.
Let us then define the infinity norm of the polynomial function defining the ground-state over $[-1,1]$ as:
\begin{align}
    \eta \coloneqq \max(|\sum_{i=0}^{D-1}c^0_iT_i(x)|, |x|\leq1).
\end{align}
We then set
\begin{align}
    \label{eq:normalized_f}
    f(x) = \frac{1}{\eta} \sum_{i=0}^{D-1}c^0_iT_i(x),
\end{align}
which satisfies the condition $|f| \leq 1$.
The coefficients $c^0_i/\eta , i=0,\dots,D-1$ are used, as described in Section~\ref{sec:qsp}, to determine the corresponding phase angles $\Phi$ of the \gls{qsp} circuit.
Let $\hat{\mathcal{U}}_\Phi$ be the unitary of the corresponding \gls{qsp} circuit,
\begin{align}
    \label{eq:u_phi}
    \hat{\mathcal{U}}_\Phi \ket{0}_a\ket{\psi_0}_s = f(\hat{H})\ket{0}_a\ket{\psi_0}_s + \beta \ket{\perp}
\end{align}
where $|\beta|^2 = 1 - \|f(\hat{H})\ket{\psi_0}_s\|^2$ and $\bra{0}_a \otimes I\ket{\perp}=0$.

The ground-state can then be re-written as 
$\ket{\Psi_0}_s = \eta \bra{0}_a \hat{\mathcal{U}}_{\Phi} \ket{0}_a\ket{\psi_0}_{s}$ 
and therefore the energy expression is $E_0=\lambda_{\mathrm{LCU}}\bra{\Psi_0}_s\hat{H}\ket{\Psi_0}_s=\lambda_{\mathrm{LCU}}\eta^2\bra{\psi_0}_s\hat{\mathcal{U}}^*_{\Phi}\ket{0}_a\bra{0}_a\hat{H}\hat{\mathcal{U}}_{\Phi}\ket{\psi_0}_s$ and therefore we have for the derivative
\begin{equation}
    \frac{\partial E_0}{\partial k_{pq}} = \lambda_{\mathrm{LCU}}\eta ^2 \bra{\psi_0}_{s} \hat{\mathcal{U}}^*_{\Phi} \ket{0}_a\bra{0}_a\hat{E}_{pq}  \hat{\mathcal{U}}_{\Phi} \ket{\psi_0}_{s} .
    \label{eq:grad_gs}
\end{equation}
Using the Jordan-Wigner mapping, the fermionic operator is replaced by a linear combination of Pauli strings, hence the derivative of Eq.~\eqref{eq:grad_gs} is ultimately a linear combination of expectation values with respect to Pauli strings $\hat{P}_{\nu}$ with coefficients $\omega_\nu$
\begin{equation}
    \frac{\partial E_0}{\partial k_{pq}} = \lambda_{\mathrm{LCU}}\eta ^2\sum_{\nu}  \omega_\nu \bra{\psi_0}_{s} \hat{\mathcal{U}}^*_{\Phi} \ket{0}_a\bra{0}_a \hat{P}_{\nu}  \hat{\mathcal{U}}_{\Phi} \ket{\psi_0}_{s}.
\end{equation}

One can now proceed in two different ways to calculate the nuclear gradients.
In the first approach, each term of the previous equation can be measured by post selecting the measurements to those where the ancilla qubits are in the zero state.
However, this has a success probability $\text{P}_{\text{suc}} \coloneqq \| f(H)\ket{\psi_0}_s\|^2$.


We therefore propose an alternative coherent, i.e. without post-selecting on the ancilla regitser, way of measuring the expectation value of $\hat{P}_{\nu}$.
This can be achieved by measuring the observables $\hat{P_\nu}$ and $\hat{R}_0\hat{P_\nu}$ of the state produced by the \gls{qsp} circuit $\hat
U_\Phi\ket{0}\ket{\phi_0}$:
\begin{equation}
\label{eq:hadamard_test_1}
    o_1 \coloneqq \bra{\psi_0} _s\bra{0}_a\hat{\mathcal{U}}_{\Phi}^{*}\hat{P}_{\nu}\hat{\mathcal{U}}_{\Phi}\ket{0}_a\ket{\psi_0}_s
\end{equation}
and
\begin{equation}
\label{eq:hadamard_test_2}
   o_2 \coloneqq \bra{\psi_0}_s\bra{0}_a\hat{\mathcal{U}}^{*}_{\Phi}\hat{R}_0 \hat{P_{\nu}}\hat{\mathcal{U}}_{\Phi}\ket{0}_a\ket{\psi_0}_s 
\end{equation}
It can then be easily demonstrated (see Appendix~\ref{app:hdmrd_test_prf}) that the sum of both observables yields:
\begin{equation}
    \label{eq:alt_approach}
    2\langle \Psi_0 |_s\hat{P}_{\nu}|\Psi_0\rangle_s = \eta^2(o_1 + o_2).
\end{equation}
It is worth noting that both $\hat{P}_\nu$ and $\hat{R}_\nu$ commute and therefore the observables $\hat{P}_\nu$ and $\hat{R}_{0}\hat{P}_\nu$ can be jointly measured. 
In the following section, we compare both approaches by benchmarking their statistical properties via numerical simulations.  


The \gls{qsp} based eigenstate preparation routine developed above opens up the possibility to combine quantum Krylov methods with other techniques for the more efficient or more accurate quantum simulation of molecules.
Instead of decomposing the Hamiltonian into Pauli words and then measuring them separately, one and two-particle \glspl{rdm} can be obtained from compressed representations of the Hamiltonian, such as different flavors of double factorization
\cite{hohenstein2022efficient,Oumarou_2024,rocca2024reducing}.
This should further reduce the variance and thus number of shots needed to compute, e.g., nuclear gradients to a desired precision.
Alternatively, one can take a classical shadow \cite{huang2020predicting,zhao_fermionic_2021,low_classical_2022,wan2022matchgate} of the state and then estimate \glspl{rdm} from that shadow.
One can even use such a shadow of a Krylov eigenstate to unbias fermionic quantum Monte-Carlo computations \cite{huggins_unbiasing_2022} or tailor split-amplitude coupled cluster methods \cite{Scheurer_2024} and thereby, e.g., augment the results with dynamic correlation corrections.
We leave a detailed investigation of the potential of these method combinations to future work and in the following restrict ourselves to a numerical comparison of the methods to compute nuclear gradients via Pauli decompositions of molecular Hamiltonians.

\section{Nuclear Gradient Benchmarks}
\label{sec:results}
\begin{figure}[b]
    \centering
    \includegraphics[width=0.5\linewidth]{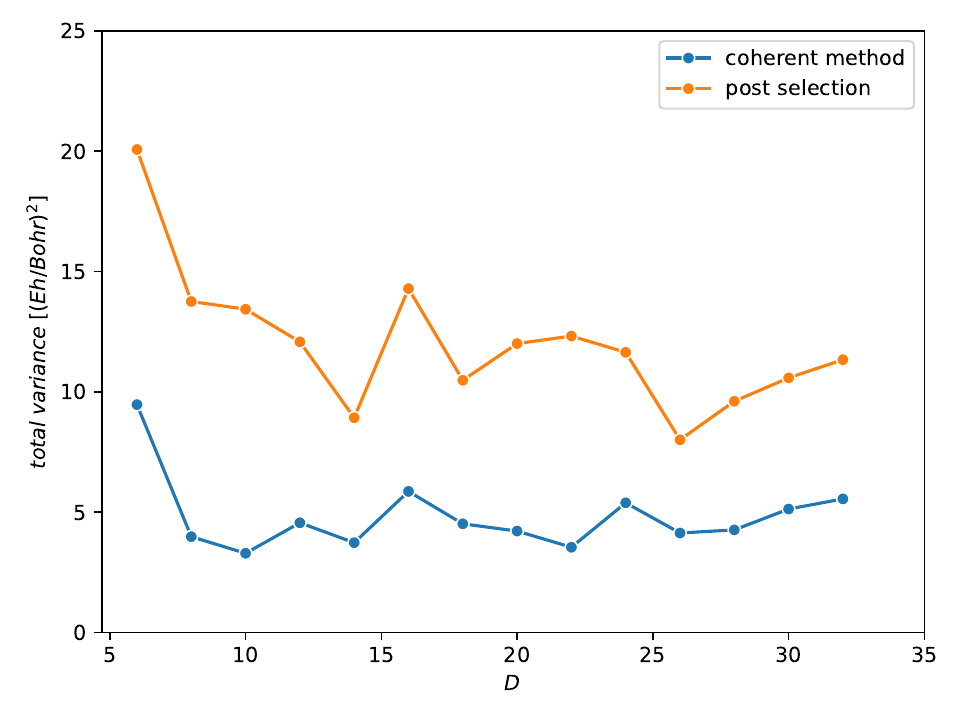}
    \caption{Total variance, i.e. sum of the variance of all the nuclear gradient $dE_0/dx$ vector components, as a function of the Krylov dimension. The blue and orange plot represent the variance of the coherent approach, that measures $\hat{P_\nu}$ and $\hat{R_{0}}\hat{P_\nu}$, and the post-selection approach respectively for $s=10^{-3}$.}
    \label{fig:variance_vs_degree}
\end{figure}

\begin{figure}
\centering 
\subfloat[Infinity norm $\eta$ as a function of the Krylov dimension $D$. The line style represents the $\eta$ values for different threshold cutoffs $s$.]{%
  \includegraphics[width=0.5\columnwidth]{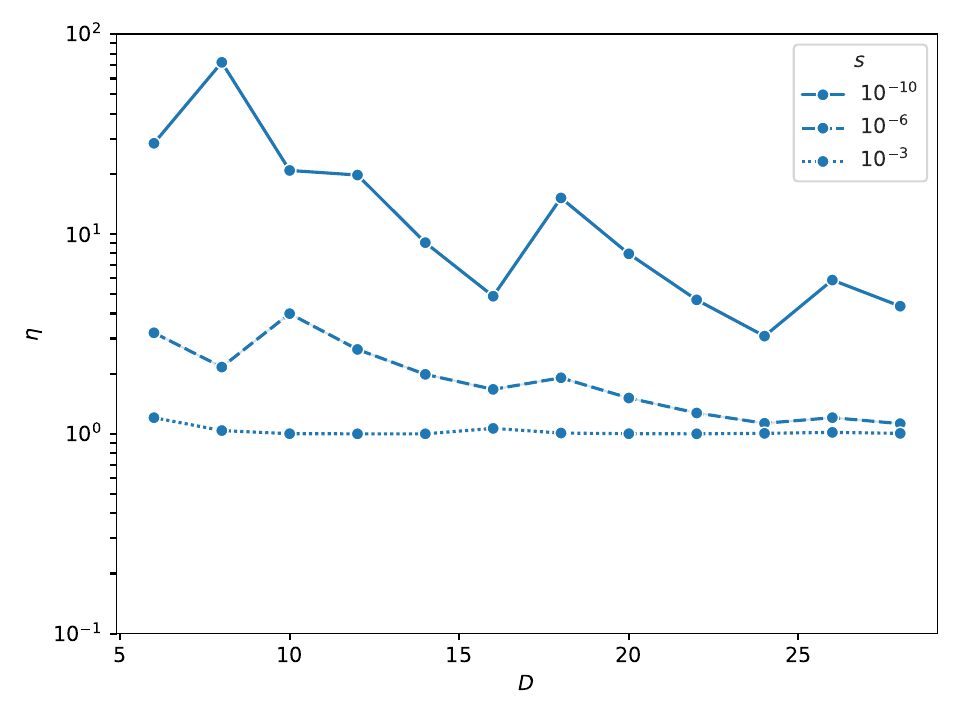}%
  \label{fig:eta_norm_vs_degree}%
}\qquad
\subfloat[Absolute difference $\Delta$ between ground-state energy and $E_0$  as a function of the Krylov dimension D. The line style represents the infinity norm $\eta$ for different threshold cutoffs $s$.]{%
  \includegraphics[width=0.5\columnwidth]{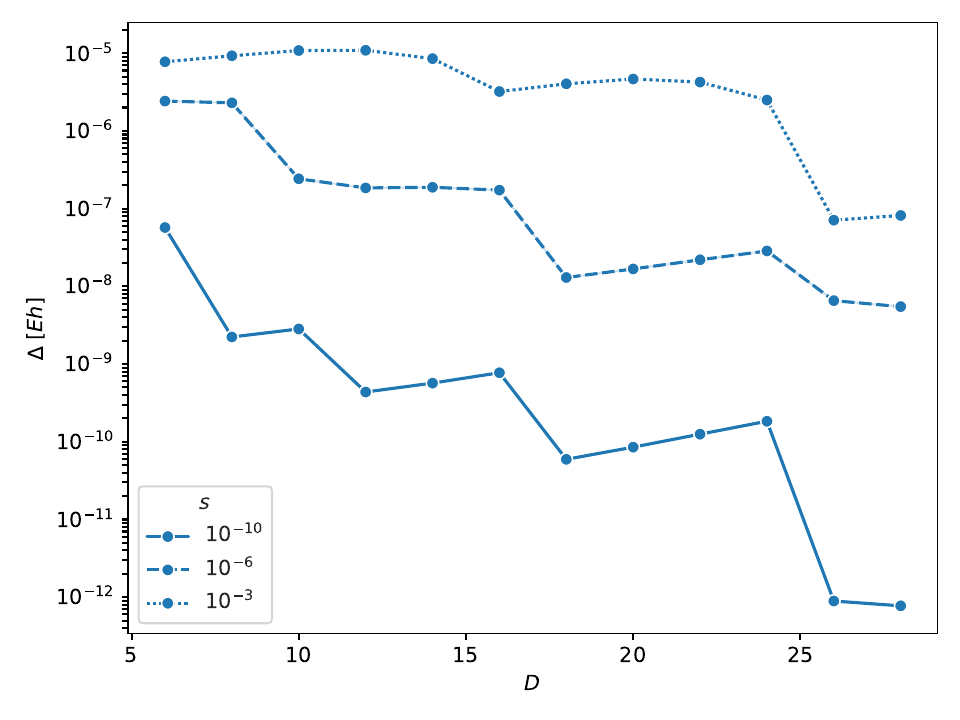}%
  \label{fig:delta_vs_degree}%
}
    \caption{FIG \ref{fig:eta_norm_vs_degree} infinity norm $\eta$ and FIG. \ref{fig:delta_vs_degree} the absolute difference of the Krylov minimal energy and the ground-state energy.}
    \label{fig:eta_end_delta}
\end{figure}
In this section we benchmark the efficiency of the nuclear gradient calculation on the quantum Krylov ground-state through numerical examples. 
To this aim, we consider the \ce{H2O} molecule in an active space of 4 electrons in 4 molecular orbitals using cc-pVDZ atomic orbital basis set. We use PySCF~\cite{sun2018pyscf} for generating the one and two-electron integrals and PennyLane~\cite{bergholm2018pennylane} to simulate the circuits. 

As per Section~\ref{sec:qsp}, we employ Chebyshev polynomials, applied to the Hartree-Fock state, to build the Krylov basis.
We simulate the measurement of the matrix elements $\widetilde{H}_{ij}\coloneqq \bra{\psi_0}T_i(\hat{H})\hat{H}T_j(\hat{H})\ket{\psi_0}$ and $\widetilde{S}_{ij} \coloneqq \bra{\psi_0}T_i(\hat{H})T_j(\hat{H})\ket{\psi_0}$ as described in section 
\MakeUppercase{\romannumeral 2} Ref.~\cite{kirby2023exact}.
The Krylov ground-state coefficients, $c^0$, are obtained by solving the generalized eigenvalue problem as described in Section~\ref{sec:background} Eq.~\eqref{eq:can_eig}.
We regularize the overlap matrix $\widetilde{S}$ by fixing a threshold, $s$ and by discarding all eigenvalues and their corresponding eigenvectors of $\widetilde{S}$ that are below $s$.


First, we compare the performance of measuring the one and two-particle \glspl{rdm} using the post-selection and the coherent approach of Eq.~\eqref{eq:alt_approach} presented in Section~\ref{sec:rdm_evaluation}. We assess how each method influences the efficiency of calculating the nuclear gradient as per Eq.~\eqref{eq:nucl_grad_chain_rule}.
To do so we first determine the analytical variance of the \glspl{rdm}' matrix elements, $\gamma_{pq}$ and $\Gamma_{pqrs}$, with PennyLane.
We then draw samples $(\gamma,\Gamma)_i$, with $1\leq i \leq 100$, where each matrix element of the \glspl{rdm} is obtained from a normal distribution defined by the analytical variance and by a mean taken to be the exact $\gamma_{pq}$ (or $\Gamma_{pqrs}$). We compute the nuclear gradient for each of the 100 realizations and compute the ensemble mean and variance.
In Figure~\ref{fig:variance_vs_degree}, we plot the total variance, i.e., the sum of the ensemble variances of all the elements of the nuclear gradient vectors for both methods (post-selection and the Eq.~\eqref{eq:alt_approach} approach).
We record the total variance's evolution as a function of the Krylov dimension, $D$.
We observe that the latter
approach results in a lower variance compared to post-selection, on average by about a factor of approximately two.
This directly implies the same factor is carried between the total shot budget of both methods for a given target precision $\epsilon$.
The optimal shot distribution should take into account the variances of both $p_0$~\eqref{eq:hadamard_test_1} and $p_1$~\eqref{eq:hadamard_test_2}.
In all cases the variance of the \gls{rdm} has a quartic dependence on the infinity norm $\eta$, i.e., 
\begin{equation}
\text{Var}[\gamma_{pq}]=\eta^4\text{Var}[\bra{\psi_0}f(\hat{H})\hat{E}_{pq}f(\hat{H})\ket{\psi_0}],
\end{equation}
and similarly for $\Gamma_{qprs}$.
It is therefore crucial to control the infinity norm $\eta$ to prevent a significant increase in the variance of the \glspl{rdm} elements, which would result in a substantially higher measurement cost.
In Figure~\ref{fig:eta_norm_vs_degree}, we plot the infinity norm $\eta$ as a function of the Krylov subspace dimension for different regularization threshold $s$.
We first observe that the value of $\eta$ generally decreases with the Krylov subspace dimension. 
More importantly, we see that setting a low threshold $s$ leads to high $\eta$, especially when $D$ is small.
For small $D$ and $s$, the increase in measurements due to $\eta$ becomes impractical, even for the small molecular system studied here.
However, enlarging $s$ has a negative effect on the energy accuracy.

In Figure~\ref{fig:delta_vs_degree}, we plot the absolute difference $\Delta$ between the true ground-state energy, obtained via exact diagonalization of the active space Hamiltonian, and the \gls{qksd} ground-state energy as a function of $D$.
This is shown for various values of $s$.
As expected, the \gls{qksd} energy converges to the exact one as $D$ increases, with smaller $s$ values providing more accurate approximations of the ground-state energy. 
In practice one thus needs to chose $s$ such that the energy error remains acceptable and then increase $D$, possibly even beyond the $D$ needed for satisfactory energy precision, to reduce $\eta$.
It will be important to study the scaling of all these quantities with the size and amount of correlation in a molecular system.

\section{Conclusion and Discussion}
In this work, we presented a method for calculating nuclear energy gradients alongside energy estimates using the \gls{qksd} approach.
We first gave the expression for the derivative of the \gls{qksd} eigenenergies with respect to any parameter of the Hamiltonian and showed that, while the energies can be obtained with a number of measurements scaling linearly with the Krylov space dimension $D$, the derivatives naively require $\mathcal{O}(D^2)$ \gls{qpu} calls.
To address this challenge, we proposed to directly prepare\gls{qksd} eigenstates $\ket{\Psi_j}$ and extract nuclear gradients by measuring its one- and two-particle \glspl{rdm}.
To efficiently obtain the coefficients of $\ket{\Psi_j}$ with respect to the Krylov basis for the use in \gls{qsp} we work with a basis of Chebyshev polynomials.
Importantly, these coefficients must be normalized by an infinity norm $\eta$ before they can be used in \gls{qsp} and the variance of the measured \gls{rdm} elements scales like $\eta^4$.
We find that mildly increasing the Krylov dimension $D$ and regularization $s$ used when solving the generalized Krylov eigenvalue problem can reduce $\eta$ significantly. 
In a post selected approach the measurement of the \glspl{rdm} elements also suffers from finite success probabilities, leading to discarded measurement outcomes and increased variance.
To mitigate this, we proposed a coherent measurement scheme utilizing two jointly measurable observables, which improves the variance of the estimator.
We demonstrated the enhanced performance of the proposed measurement technique through numerical simulations.

Our numerical results highlight the impact of the generalized eigenvalue regularization threshold $s$ on both the energy approximation and the variance of the \glspl{rdm} elements.
While tight thresholds yield a better energy precision, they also lead to larger $\eta$ and therefore increase the measurement cost of the \glspl{rdm}.
It is therefore important to balance this parameter in \gls{qksd} calculations that are aimed at more than just computing energies. 
In practice, $s$ is often chosen to scale linearly with the noise level.
Ref.~\cite{Lee_2024} demonstrated that in presence of sampling noise, i.e., a finite number of measurements $M$ for each matrix element $\widetilde{H}_{ij}$ and $\widetilde{S}_{ij}$, the optimal value of $s$ is $s_{\text{opt}} = e / \sqrt{M}$ where $e$ is a factor that depends linearly on the dimension of the Hilbert space. 
In future work, it would be interesting to explore how, through this choice of $s$, the noise level affects $\eta$ and therefore the efficiency of calculating properties in our \gls{qksd} and \gls{qsp} framework. 

Obtaining the \glspl{rdm} via our coherent approach reduces the measurement cost per matrix element from $\mathcal{O}(D^2)$ to $\mathcal{O}(1)$ compared to the post-selection approach.
It is important to realize that this is achieved without increasing circuit depth.
This is because both approaches rely on \gls{qsp} to implement the Chebyshev polynomials, and the depth of the \gls{qsp} circuit is determined by the polynomial degree, which is $D$ in both cases.


In this manuscript, we have focused on the nuclear gradients.
However, the techniques developed are broadly applicable to any property that requires the measurement of the one- and/or two-particle \glspl{rdm} including dipole moments or natural orbitals.
In addition, our procedure can be used to compute excited state \glspl{rdm} and gradients if a suitable reference state is used.

\section{Acknowledgments}
The authors thank J\'er\^ome Gonthier for useful discussions and feedback on the manuscript.
P.J.O.\ and R.M.P.\ own stock/options in QC Ware Corp.

\bibliography{main}

\clearpage
\begin{appendices}
    \input{appendix}
\end{appendices}

\end{document}

%% file: appendix.tex
\onecolumngrid

\section{Derivative of Krylov Energy}
\label{app:proofs}
Let $(E_m,\ket{\xi_m})$ be a generalized eigenpair w.r.t $(\widetilde{H}, \widetilde{S})$ i.e. 
\begin{equation}
\label{eq:app1}
    \widetilde{H}\ket{\xi_m}=E_m\widetilde{S}\ket{\xi_m}
\end{equation} which subsequently implies that $\bra{\xi_m}\widetilde{H} \ket{\xi_m}=E_m\bra{\xi_m}\widetilde{S}\ket{\xi_m}$
\begin{align}
    \frac{d \bra{\xi_m}\widetilde{H} \ket{\xi_m}}{dx}&=\frac{dE_m}{dx} \bra{\xi_m}\widetilde{S}\ket{\xi_m} + E_m\frac{ d\bra{\xi_m}\widetilde{S}\ket{\xi_m}}{dx}\\
    \bra{\frac{d\xi_m}{dx}}\widetilde{H} \ket{\xi_m} + \bra{\xi_m} \frac{d\widetilde{H}}{dx} \ket{\xi_m} + \bra{\xi_m}\widetilde{H} \ket{\frac{d\xi_m}{dx}}&=\frac{dE_m}{dx} \bra{\xi_m}\widetilde{S}\ket{\xi_m} + E_m (\bra{\frac{d\xi_m}{dx}}\widetilde{S} \ket{\xi_m} + \bra{\xi_m} \frac{d\widetilde{S}}{dx} \ket{\xi_m} + \bra{\xi_m}\widetilde{S} \ket{\frac{d\xi_m}{dx}}) 
\end{align}
using \eqref{eq:app1} we have:
\begin{align}
    E_m\bra{\frac{d\xi_m}{dx}}\widetilde{S} \ket{\xi_m} + \bra{\xi_m} \frac{d\widetilde{H}}{dx} \ket{\xi_m} + E_m\bra{\xi_m}\widetilde{S} \ket{\frac{d\xi_m}{dx}}&=\frac{dE_m}{dx} \bra{\xi_m}\widetilde{S}\ket{\xi_m} + E_m (\bra{\frac{d\xi_m}{dx}}\widetilde{S} \ket{\xi_m} + \bra{\xi_m} \frac{d\widetilde{S}}{dx} \ket{\xi_m} + \bra{\xi_m}\widetilde{S} \ket{\frac{d\xi_m}{dx}}) \\
    \implies \bra{\xi_m} \frac{d\widetilde{H}}{dx} \ket{\xi_m} &= \frac{dE_m}{dx} \bra{\xi_m}\widetilde{S}\ket{\xi_m} + E_m\bra{\xi_m} \frac{d\widetilde{S}}{dx}\ket{\xi_m}\\
    \implies \frac{dE_m}{dx} &= \frac{\bra{\xi_m} \frac{d\widetilde{H}}{dx} \ket{\xi_m}\bra{\xi_m}\widetilde{S}\ket{\xi_m}-\bra{\xi_m} \frac{d\widetilde{S}}{dx} \ket{\xi_m}\bra{\xi_m}\widetilde{H}\ket{\xi_m}}{\bra{\xi_m}\widetilde{S}\ket{\xi_m}^2}  
\end{align}

\section{\gls{rdm} through a coherent approach}
\label{app:hdmrd_test_prf}
Let us first recall the function $f$ and unitary $\hat{\mathcal{U}}_\Phi$ defined in ~\eqref{eq:normalized_f} and ~\eqref{eq:u_phi} respectively. Let $\hat{P}_\nu$ and $\hat{R}_{0}$ be a Pauli string and the reflection defined in \eqref{eq:reflection}. Then picking up from equation \eqref{eq:u_phi}, we have:
\begin{align}
    & \hat{\mathcal{U}}_\Phi\ket{0}_a \ket{\phi_0}_s = f(\hat{H})\ket{0}_a\ket{\phi_0}_s + \beta \ket{\perp}\\
    \implies & \bra{\phi_0}_s\bra{0}_a\hat{\mathcal{U}}^*_\Phi  \hat{P}_\nu \hat{\mathcal{U}}_\Phi\ket{0}_a \ket{\phi_0}_s = \frac{1}{\eta^2}\bra{\Psi_0}\hat{P}_\nu\ket{\Psi_0} + \beta^2 \bra{\perp}\hat{P}_\nu\ket{\perp}, \label{pr_eq:1}
\end{align}
and we have
\begin{align}
     & \hat{R}_{0}\hat{P}_\nu \hat{\mathcal{U}}_\Phi\ket{0}_a \ket{\phi_0}_s = (2\ket{0}\bra{0}_a-I)\hat{P}_\nu f(\hat{H})\ket{0}_a\ket{\phi_0} + \beta (2\ket{0}\bra{0}_a-I)\hat{P}_\nu\ket{\perp}
     = \hat{P}_\nu f(\hat{H})\ket{0}_a\ket{\phi_0} - \beta \ket{\perp} \\
     \implies & \bra{\phi_0}_s\bra{0}_a\hat{\mathcal{U}}^*_\Phi  \hat{R}_{0}\hat{P}_\nu \hat{\mathcal{U}}_\Phi\ket{0}_a \ket{\phi_0}_s=\bra{\phi_0}f(\hat{H})\hat{P}_\nu f(\hat{H})\ket{\phi_0} - \beta^2\bra{\perp}\hat{P}_\nu \ket{\perp}=\frac{1}{\eta^2}\bra{\Psi_0}\hat{P}_\nu\ket{\Psi_0} - \beta^2\bra{\perp}\hat{P}_\nu\ket{\perp} \label{pr_eq:2}
\end{align}
Adding \eqref{pr_eq:1} and \eqref{pr_eq:2} we obtain
\begin{align}
    \bra{\phi_0}_s\bra{0}_a\hat{\mathcal{U}}^*_\Phi  \hat{P}_\nu \hat{\mathcal{U}}_\Phi\ket{0}_a \ket{\phi_0}_s + \bra{\phi_0}_s\bra{0}_a\hat{\mathcal{U}}^*_\Phi  \hat{R}_{0}\hat{P}_\nu \hat{\mathcal{U}}_\Phi\ket{0}_a \ket{\phi_0}_s=\bra{\phi_0}f(\hat{H})\hat{P}_\nu f(\hat{H})\ket{\phi_0} - \beta^2\bra{\perp}\hat{P}_\nu \ket{\perp}
    = \frac{2}{\eta^2}\bra{\Psi_0}\hat{P_{\nu}}\ket{\Psi_0}.
\end{align}

%% file: main.bbl
\begin{thebibliography}{70}%
\makeatletter
\providecommand \@ifxundefined [1]{%
 \@ifx{#1\undefined}
}%
\providecommand \@ifnum [1]{%
 \ifnum #1\expandafter \@firstoftwo
 \else \expandafter \@secondoftwo
 \fi
}%
\providecommand \@ifx [1]{%
 \ifx #1\expandafter \@firstoftwo
 \else \expandafter \@secondoftwo
 \fi
}%
\providecommand \natexlab [1]{#1}%
\providecommand \enquote  [1]{``#1''}%
\providecommand \bibnamefont  [1]{#1}%
\providecommand \bibfnamefont [1]{#1}%
\providecommand \citenamefont [1]{#1}%
\providecommand \href@noop [0]{\@secondoftwo}%
\providecommand \href [0]{\begingroup \@sanitize@url \@href}%
\providecommand \@href[1]{\@@startlink{#1}\@@href}%
\providecommand \@@href[1]{\endgroup#1\@@endlink}%
\providecommand \@sanitize@url [0]{\catcode `\\12\catcode `\$12\catcode `\&12\catcode `\#12\catcode `\^12\catcode `\_12\catcode `\%12\relax}%
\providecommand \@@startlink[1]{}%
\providecommand \@@endlink[0]{}%
\providecommand \url  [0]{\begingroup\@sanitize@url \@url }%
\providecommand \@url [1]{\endgroup\@href {#1}{\urlprefix }}%
\providecommand \urlprefix  [0]{URL }%
\providecommand \Eprint [0]{\href }%
\providecommand \doibase [0]{https://doi.org/}%
\providecommand \selectlanguage [0]{\@gobble}%
\providecommand \bibinfo  [0]{\@secondoftwo}%
\providecommand \bibfield  [0]{\@secondoftwo}%
\providecommand \translation [1]{[#1]}%
\providecommand \BibitemOpen [0]{}%
\providecommand \bibitemStop [0]{}%
\providecommand \bibitemNoStop [0]{.\EOS\space}%
\providecommand \EOS [0]{\spacefactor3000\relax}%
\providecommand \BibitemShut  [1]{\csname bibitem#1\endcsname}%
\let\auto@bib@innerbib\@empty
\bibitem [{\citenamefont {Reiher}\ \emph {et~al.}(2017)\citenamefont {Reiher}, \citenamefont {Wiebe}, \citenamefont {Svore}, \citenamefont {Wecker},\ and\ \citenamefont {Troyer}}]{reiher2017elucidating}%
  \BibitemOpen
  \bibfield  {author} {\bibinfo {author} {\bibfnamefont {M.}~\bibnamefont {Reiher}}, \bibinfo {author} {\bibfnamefont {N.}~\bibnamefont {Wiebe}}, \bibinfo {author} {\bibfnamefont {K.~M.}\ \bibnamefont {Svore}}, \bibinfo {author} {\bibfnamefont {D.}~\bibnamefont {Wecker}},\ and\ \bibinfo {author} {\bibfnamefont {M.}~\bibnamefont {Troyer}},\ }\bibfield  {title} {\bibinfo {title} {Elucidating reaction mechanisms on quantum computers},\ }\href@noop {} {\bibfield  {journal} {\bibinfo  {journal} {Proceedings of the national academy of sciences}\ }\textbf {\bibinfo {volume} {114}},\ \bibinfo {pages} {7555} (\bibinfo {year} {2017})}\BibitemShut {NoStop}%
\bibitem [{\citenamefont {Babbush}\ \emph {et~al.}(2018)\citenamefont {Babbush}, \citenamefont {Gidney}, \citenamefont {Berry}, \citenamefont {Wiebe}, \citenamefont {McClean}, \citenamefont {Paler}, \citenamefont {Fowler},\ and\ \citenamefont {Neven}}]{babbush2018encoding}%
  \BibitemOpen
  \bibfield  {author} {\bibinfo {author} {\bibfnamefont {R.}~\bibnamefont {Babbush}}, \bibinfo {author} {\bibfnamefont {C.}~\bibnamefont {Gidney}}, \bibinfo {author} {\bibfnamefont {D.~W.}\ \bibnamefont {Berry}}, \bibinfo {author} {\bibfnamefont {N.}~\bibnamefont {Wiebe}}, \bibinfo {author} {\bibfnamefont {J.}~\bibnamefont {McClean}}, \bibinfo {author} {\bibfnamefont {A.}~\bibnamefont {Paler}}, \bibinfo {author} {\bibfnamefont {A.}~\bibnamefont {Fowler}},\ and\ \bibinfo {author} {\bibfnamefont {H.}~\bibnamefont {Neven}},\ }\bibfield  {title} {\bibinfo {title} {Encoding electronic spectra in quantum circuits with linear t complexity},\ }\href@noop {} {\bibfield  {journal} {\bibinfo  {journal} {Physical Review X}\ }\textbf {\bibinfo {volume} {8}},\ \bibinfo {pages} {041015} (\bibinfo {year} {2018})}\BibitemShut {NoStop}%
\bibitem [{\citenamefont {Berry}\ \emph {et~al.}(2019)\citenamefont {Berry}, \citenamefont {Gidney}, \citenamefont {Motta}, \citenamefont {McClean},\ and\ \citenamefont {Babbush}}]{berry2019qubitization}%
  \BibitemOpen
  \bibfield  {author} {\bibinfo {author} {\bibfnamefont {D.~W.}\ \bibnamefont {Berry}}, \bibinfo {author} {\bibfnamefont {C.}~\bibnamefont {Gidney}}, \bibinfo {author} {\bibfnamefont {M.}~\bibnamefont {Motta}}, \bibinfo {author} {\bibfnamefont {J.~R.}\ \bibnamefont {McClean}},\ and\ \bibinfo {author} {\bibfnamefont {R.}~\bibnamefont {Babbush}},\ }\bibfield  {title} {\bibinfo {title} {Qubitization of arbitrary basis quantum chemistry leveraging sparsity and low rank factorization},\ }\href@noop {} {\bibfield  {journal} {\bibinfo  {journal} {Quantum}\ }\textbf {\bibinfo {volume} {3}},\ \bibinfo {pages} {208} (\bibinfo {year} {2019})}\BibitemShut {NoStop}%
\bibitem [{\citenamefont {Lee}\ \emph {et~al.}(2021)\citenamefont {Lee}, \citenamefont {Berry}, \citenamefont {Gidney}, \citenamefont {Huggins}, \citenamefont {McClean}, \citenamefont {Wiebe},\ and\ \citenamefont {Babbush}}]{lee2021even}%
  \BibitemOpen
  \bibfield  {author} {\bibinfo {author} {\bibfnamefont {J.}~\bibnamefont {Lee}}, \bibinfo {author} {\bibfnamefont {D.~W.}\ \bibnamefont {Berry}}, \bibinfo {author} {\bibfnamefont {C.}~\bibnamefont {Gidney}}, \bibinfo {author} {\bibfnamefont {W.~J.}\ \bibnamefont {Huggins}}, \bibinfo {author} {\bibfnamefont {J.~R.}\ \bibnamefont {McClean}}, \bibinfo {author} {\bibfnamefont {N.}~\bibnamefont {Wiebe}},\ and\ \bibinfo {author} {\bibfnamefont {R.}~\bibnamefont {Babbush}},\ }\bibfield  {title} {\bibinfo {title} {Even more efficient quantum computations of chemistry through tensor hypercontraction},\ }\href@noop {} {\bibfield  {journal} {\bibinfo  {journal} {PRX Quantum}\ }\textbf {\bibinfo {volume} {2}},\ \bibinfo {pages} {030305} (\bibinfo {year} {2021})}\BibitemShut {NoStop}%
\bibitem [{\citenamefont {Su}\ \emph {et~al.}(2021)\citenamefont {Su}, \citenamefont {Berry}, \citenamefont {Wiebe}, \citenamefont {Rubin},\ and\ \citenamefont {Babbush}}]{su2021fault}%
  \BibitemOpen
  \bibfield  {author} {\bibinfo {author} {\bibfnamefont {Y.}~\bibnamefont {Su}}, \bibinfo {author} {\bibfnamefont {D.~W.}\ \bibnamefont {Berry}}, \bibinfo {author} {\bibfnamefont {N.}~\bibnamefont {Wiebe}}, \bibinfo {author} {\bibfnamefont {N.}~\bibnamefont {Rubin}},\ and\ \bibinfo {author} {\bibfnamefont {R.}~\bibnamefont {Babbush}},\ }\bibfield  {title} {\bibinfo {title} {Fault-tolerant quantum simulations of chemistry in first quantization},\ }\href@noop {} {\bibfield  {journal} {\bibinfo  {journal} {PRX Quantum}\ }\textbf {\bibinfo {volume} {2}},\ \bibinfo {pages} {040332} (\bibinfo {year} {2021})}\BibitemShut {NoStop}%
\bibitem [{\citenamefont {Kim}\ \emph {et~al.}(2022)\citenamefont {Kim}, \citenamefont {Liu}, \citenamefont {Pallister}, \citenamefont {Pol}, \citenamefont {Roberts},\ and\ \citenamefont {Lee}}]{kim2022fault}%
  \BibitemOpen
  \bibfield  {author} {\bibinfo {author} {\bibfnamefont {I.~H.}\ \bibnamefont {Kim}}, \bibinfo {author} {\bibfnamefont {Y.-H.}\ \bibnamefont {Liu}}, \bibinfo {author} {\bibfnamefont {S.}~\bibnamefont {Pallister}}, \bibinfo {author} {\bibfnamefont {W.}~\bibnamefont {Pol}}, \bibinfo {author} {\bibfnamefont {S.}~\bibnamefont {Roberts}},\ and\ \bibinfo {author} {\bibfnamefont {E.}~\bibnamefont {Lee}},\ }\bibfield  {title} {\bibinfo {title} {Fault-tolerant resource estimate for quantum chemical simulations: Case study on li-ion battery electrolyte molecules},\ }\href@noop {} {\bibfield  {journal} {\bibinfo  {journal} {Physical Review Research}\ }\textbf {\bibinfo {volume} {4}},\ \bibinfo {pages} {023019} (\bibinfo {year} {2022})}\BibitemShut {NoStop}%
\bibitem [{\citenamefont {Delgado}\ \emph {et~al.}(2022)\citenamefont {Delgado}, \citenamefont {Casares}, \citenamefont {Dos~Reis}, \citenamefont {Zini}, \citenamefont {Campos}, \citenamefont {Cruz-Hern{\'a}ndez}, \citenamefont {Voigt}, \citenamefont {Lowe}, \citenamefont {Jahangiri}, \citenamefont {Martin-Delgado} \emph {et~al.}}]{delgado2022simulating}%
  \BibitemOpen
  \bibfield  {author} {\bibinfo {author} {\bibfnamefont {A.}~\bibnamefont {Delgado}}, \bibinfo {author} {\bibfnamefont {P.~A.}\ \bibnamefont {Casares}}, \bibinfo {author} {\bibfnamefont {R.}~\bibnamefont {Dos~Reis}}, \bibinfo {author} {\bibfnamefont {M.~S.}\ \bibnamefont {Zini}}, \bibinfo {author} {\bibfnamefont {R.}~\bibnamefont {Campos}}, \bibinfo {author} {\bibfnamefont {N.}~\bibnamefont {Cruz-Hern{\'a}ndez}}, \bibinfo {author} {\bibfnamefont {A.-C.}\ \bibnamefont {Voigt}}, \bibinfo {author} {\bibfnamefont {A.}~\bibnamefont {Lowe}}, \bibinfo {author} {\bibfnamefont {S.}~\bibnamefont {Jahangiri}}, \bibinfo {author} {\bibfnamefont {M.~A.}\ \bibnamefont {Martin-Delgado}}, \emph {et~al.},\ }\bibfield  {title} {\bibinfo {title} {Simulating key properties of lithium-ion batteries with a fault-tolerant quantum computer},\ }\href@noop {} {\bibfield  {journal} {\bibinfo  {journal} {Physical Review A}\ }\textbf {\bibinfo {volume} {106}},\ \bibinfo {pages} {032428} (\bibinfo {year} {2022})}\BibitemShut {NoStop}%
\bibitem [{\citenamefont {von Burg}\ \emph {et~al.}(2021)\citenamefont {von Burg}, \citenamefont {Low}, \citenamefont {H{\"a}ner}, \citenamefont {Steiger}, \citenamefont {Reiher}, \citenamefont {Roetteler},\ and\ \citenamefont {Troyer}}]{von2021quantum}%
  \BibitemOpen
  \bibfield  {author} {\bibinfo {author} {\bibfnamefont {V.}~\bibnamefont {von Burg}}, \bibinfo {author} {\bibfnamefont {G.~H.}\ \bibnamefont {Low}}, \bibinfo {author} {\bibfnamefont {T.}~\bibnamefont {H{\"a}ner}}, \bibinfo {author} {\bibfnamefont {D.~S.}\ \bibnamefont {Steiger}}, \bibinfo {author} {\bibfnamefont {M.}~\bibnamefont {Reiher}}, \bibinfo {author} {\bibfnamefont {M.}~\bibnamefont {Roetteler}},\ and\ \bibinfo {author} {\bibfnamefont {M.}~\bibnamefont {Troyer}},\ }\bibfield  {title} {\bibinfo {title} {Quantum computing enhanced computational catalysis},\ }\href@noop {} {\bibfield  {journal} {\bibinfo  {journal} {Physical Review Research}\ }\textbf {\bibinfo {volume} {3}},\ \bibinfo {pages} {033055} (\bibinfo {year} {2021})}\BibitemShut {NoStop}%
\bibitem [{\citenamefont {Goings}\ \emph {et~al.}(2022)\citenamefont {Goings}, \citenamefont {White}, \citenamefont {Lee}, \citenamefont {Tautermann}, \citenamefont {Degroote}, \citenamefont {Gidney}, \citenamefont {Shiozaki}, \citenamefont {Babbush},\ and\ \citenamefont {Rubin}}]{goings2022reliably}%
  \BibitemOpen
  \bibfield  {author} {\bibinfo {author} {\bibfnamefont {J.~J.}\ \bibnamefont {Goings}}, \bibinfo {author} {\bibfnamefont {A.}~\bibnamefont {White}}, \bibinfo {author} {\bibfnamefont {J.}~\bibnamefont {Lee}}, \bibinfo {author} {\bibfnamefont {C.~S.}\ \bibnamefont {Tautermann}}, \bibinfo {author} {\bibfnamefont {M.}~\bibnamefont {Degroote}}, \bibinfo {author} {\bibfnamefont {C.}~\bibnamefont {Gidney}}, \bibinfo {author} {\bibfnamefont {T.}~\bibnamefont {Shiozaki}}, \bibinfo {author} {\bibfnamefont {R.}~\bibnamefont {Babbush}},\ and\ \bibinfo {author} {\bibfnamefont {N.~C.}\ \bibnamefont {Rubin}},\ }\bibfield  {title} {\bibinfo {title} {Reliably assessing the electronic structure of cytochrome p450 on today’s classical computers and tomorrow’s quantum computers},\ }\href@noop {} {\bibfield  {journal} {\bibinfo  {journal} {Proceedings of the National Academy of Sciences}\ }\textbf {\bibinfo {volume} {119}},\ \bibinfo {pages} {e2203533119} (\bibinfo {year} {2022})}\BibitemShut {NoStop}%
\bibitem [{\citenamefont {Rocca}\ \emph {et~al.}(2024)\citenamefont {Rocca}, \citenamefont {Cortes}, \citenamefont {Gonthier}, \citenamefont {Ollitrault}, \citenamefont {Parrish}, \citenamefont {Anselmetti}, \citenamefont {Degroote}, \citenamefont {Moll}, \citenamefont {Santagati},\ and\ \citenamefont {Streif}}]{rocca2024reducing}%
  \BibitemOpen
  \bibfield  {author} {\bibinfo {author} {\bibfnamefont {D.}~\bibnamefont {Rocca}}, \bibinfo {author} {\bibfnamefont {C.~L.}\ \bibnamefont {Cortes}}, \bibinfo {author} {\bibfnamefont {J.~F.}\ \bibnamefont {Gonthier}}, \bibinfo {author} {\bibfnamefont {P.~J.}\ \bibnamefont {Ollitrault}}, \bibinfo {author} {\bibfnamefont {R.~M.}\ \bibnamefont {Parrish}}, \bibinfo {author} {\bibfnamefont {G.-L.}\ \bibnamefont {Anselmetti}}, \bibinfo {author} {\bibfnamefont {M.}~\bibnamefont {Degroote}}, \bibinfo {author} {\bibfnamefont {N.}~\bibnamefont {Moll}}, \bibinfo {author} {\bibfnamefont {R.}~\bibnamefont {Santagati}},\ and\ \bibinfo {author} {\bibfnamefont {M.}~\bibnamefont {Streif}},\ }\bibfield  {title} {\bibinfo {title} {Reducing the runtime of fault-tolerant quantum simulations in chemistry through symmetry-compressed double factorization},\ }\href@noop {} {\bibfield  {journal} {\bibinfo  {journal} {Journal of Chemical Theory and Computation}\ } (\bibinfo {year} {2024})}\BibitemShut {NoStop}%
\bibitem [{\citenamefont {Lee}\ \emph {et~al.}(2023)\citenamefont {Lee}, \citenamefont {Lee}, \citenamefont {Zhai}, \citenamefont {Tong}, \citenamefont {Dalzell}, \citenamefont {Kumar}, \citenamefont {Helms}, \citenamefont {Gray}, \citenamefont {Cui}, \citenamefont {Liu} \emph {et~al.}}]{lee2023evaluating}%
  \BibitemOpen
  \bibfield  {author} {\bibinfo {author} {\bibfnamefont {S.}~\bibnamefont {Lee}}, \bibinfo {author} {\bibfnamefont {J.}~\bibnamefont {Lee}}, \bibinfo {author} {\bibfnamefont {H.}~\bibnamefont {Zhai}}, \bibinfo {author} {\bibfnamefont {Y.}~\bibnamefont {Tong}}, \bibinfo {author} {\bibfnamefont {A.~M.}\ \bibnamefont {Dalzell}}, \bibinfo {author} {\bibfnamefont {A.}~\bibnamefont {Kumar}}, \bibinfo {author} {\bibfnamefont {P.}~\bibnamefont {Helms}}, \bibinfo {author} {\bibfnamefont {J.}~\bibnamefont {Gray}}, \bibinfo {author} {\bibfnamefont {Z.-H.}\ \bibnamefont {Cui}}, \bibinfo {author} {\bibfnamefont {W.}~\bibnamefont {Liu}}, \emph {et~al.},\ }\bibfield  {title} {\bibinfo {title} {Evaluating the evidence for exponential quantum advantage in ground-state quantum chemistry},\ }\href {https://doi.org/https://doi.org/10.1038/s41467-023-37587-6} {\bibfield  {journal} {\bibinfo  {journal} {Nature Communications}\ }\textbf {\bibinfo {volume} {14}},\ \bibinfo {pages} {1952} (\bibinfo {year} {2023})}\BibitemShut
  {NoStop}%
\bibitem [{\citenamefont {Marti-Dafcik}\ \emph {et~al.}(2024)\citenamefont {Marti-Dafcik}, \citenamefont {Lee}, \citenamefont {Burton},\ and\ \citenamefont {Tew}}]{marti2024spin}%
  \BibitemOpen
  \bibfield  {author} {\bibinfo {author} {\bibfnamefont {D.}~\bibnamefont {Marti-Dafcik}}, \bibinfo {author} {\bibfnamefont {N.}~\bibnamefont {Lee}}, \bibinfo {author} {\bibfnamefont {H.~G.}\ \bibnamefont {Burton}},\ and\ \bibinfo {author} {\bibfnamefont {D.~P.}\ \bibnamefont {Tew}},\ }\bibfield  {title} {\bibinfo {title} {Spin-coupled molecular orbitals: chemical intuition meets quantum chemistry},\ }\bibfield  {journal} {\bibinfo  {journal} {arXiv preprint arXiv:2402.08858}\ }\href {https://doi.org/https://doi.org/10.48550/arXiv.2402.08858} {https://doi.org/10.48550/arXiv.2402.08858} (\bibinfo {year} {2024})\BibitemShut {NoStop}%
\bibitem [{\citenamefont {Ollitrault}\ \emph {et~al.}(2024)\citenamefont {Ollitrault}, \citenamefont {Cortes}, \citenamefont {Gonthier}, \citenamefont {Parrish}, \citenamefont {Rocca}, \citenamefont {Anselmetti}, \citenamefont {Degroote}, \citenamefont {Moll}, \citenamefont {Santagati},\ and\ \citenamefont {Streif}}]{ollitrault2024enhancing}%
  \BibitemOpen
  \bibfield  {author} {\bibinfo {author} {\bibfnamefont {P.~J.}\ \bibnamefont {Ollitrault}}, \bibinfo {author} {\bibfnamefont {C.~L.}\ \bibnamefont {Cortes}}, \bibinfo {author} {\bibfnamefont {J.~F.}\ \bibnamefont {Gonthier}}, \bibinfo {author} {\bibfnamefont {R.~M.}\ \bibnamefont {Parrish}}, \bibinfo {author} {\bibfnamefont {D.}~\bibnamefont {Rocca}}, \bibinfo {author} {\bibfnamefont {G.-L.}\ \bibnamefont {Anselmetti}}, \bibinfo {author} {\bibfnamefont {M.}~\bibnamefont {Degroote}}, \bibinfo {author} {\bibfnamefont {N.}~\bibnamefont {Moll}}, \bibinfo {author} {\bibfnamefont {R.}~\bibnamefont {Santagati}},\ and\ \bibinfo {author} {\bibfnamefont {M.}~\bibnamefont {Streif}},\ }\bibfield  {title} {\bibinfo {title} {Enhancing initial state overlap through orbital optimization for faster molecular electronic ground-state energy estimation},\ }\href@noop {} {\bibfield  {journal} {\bibinfo  {journal} {arXiv preprint arXiv:2404.08565}\ } (\bibinfo {year} {2024})}\BibitemShut {NoStop}%
\bibitem [{\citenamefont {M{\"o}rchen}\ \emph {et~al.}(2024)\citenamefont {M{\"o}rchen}, \citenamefont {Low}, \citenamefont {Weymuth}, \citenamefont {Liu}, \citenamefont {Troyer},\ and\ \citenamefont {Reiher}}]{morchen2024classification}%
  \BibitemOpen
  \bibfield  {author} {\bibinfo {author} {\bibfnamefont {M.}~\bibnamefont {M{\"o}rchen}}, \bibinfo {author} {\bibfnamefont {G.~H.}\ \bibnamefont {Low}}, \bibinfo {author} {\bibfnamefont {T.}~\bibnamefont {Weymuth}}, \bibinfo {author} {\bibfnamefont {H.}~\bibnamefont {Liu}}, \bibinfo {author} {\bibfnamefont {M.}~\bibnamefont {Troyer}},\ and\ \bibinfo {author} {\bibfnamefont {M.}~\bibnamefont {Reiher}},\ }\bibfield  {title} {\bibinfo {title} {Classification of electronic structures and state preparation for quantum computation of reaction chemistry},\ }\href@noop {} {\bibfield  {journal} {\bibinfo  {journal} {arXiv preprint arXiv:2409.08910}\ } (\bibinfo {year} {2024})}\BibitemShut {NoStop}%
\bibitem [{\citenamefont {Berry}\ \emph {et~al.}(2015)\citenamefont {Berry}, \citenamefont {Childs},\ and\ \citenamefont {Kothari}}]{berry2015hamiltonian}%
  \BibitemOpen
  \bibfield  {author} {\bibinfo {author} {\bibfnamefont {D.~W.}\ \bibnamefont {Berry}}, \bibinfo {author} {\bibfnamefont {A.~M.}\ \bibnamefont {Childs}},\ and\ \bibinfo {author} {\bibfnamefont {R.}~\bibnamefont {Kothari}},\ }\bibfield  {title} {\bibinfo {title} {Hamiltonian simulation with nearly optimal dependence on all parameters},\ }in\ \href@noop {} {\emph {\bibinfo {booktitle} {2015 IEEE 56th annual symposium on foundations of computer science}}}\ (\bibinfo {organization} {IEEE},\ \bibinfo {year} {2015})\ pp.\ \bibinfo {pages} {792--809}\BibitemShut {NoStop}%
\bibitem [{\citenamefont {Low}\ and\ \citenamefont {Chuang}(2017)}]{low2017optimal}%
  \BibitemOpen
  \bibfield  {author} {\bibinfo {author} {\bibfnamefont {G.~H.}\ \bibnamefont {Low}}\ and\ \bibinfo {author} {\bibfnamefont {I.~L.}\ \bibnamefont {Chuang}},\ }\bibfield  {title} {\bibinfo {title} {Optimal hamiltonian simulation by quantum signal processing},\ }\href@noop {} {\bibfield  {journal} {\bibinfo  {journal} {Physical review letters}\ }\textbf {\bibinfo {volume} {118}},\ \bibinfo {pages} {010501} (\bibinfo {year} {2017})}\BibitemShut {NoStop}%
\bibitem [{\citenamefont {Low}\ and\ \citenamefont {Chuang}(2019)}]{low2019hamiltonian}%
  \BibitemOpen
  \bibfield  {author} {\bibinfo {author} {\bibfnamefont {G.~H.}\ \bibnamefont {Low}}\ and\ \bibinfo {author} {\bibfnamefont {I.~L.}\ \bibnamefont {Chuang}},\ }\bibfield  {title} {\bibinfo {title} {Hamiltonian simulation by qubitization},\ }\href {https://doi.org/10.22331/q-2019-07-12-163} {\bibfield  {journal} {\bibinfo  {journal} {Quantum}\ }\textbf {\bibinfo {volume} {3}},\ \bibinfo {pages} {163} (\bibinfo {year} {2019})}\BibitemShut {NoStop}%
\bibitem [{\citenamefont {Griffiths}\ and\ \citenamefont {Niu}(1996)}]{griffiths1996semiclassical}%
  \BibitemOpen
  \bibfield  {author} {\bibinfo {author} {\bibfnamefont {R.~B.}\ \bibnamefont {Griffiths}}\ and\ \bibinfo {author} {\bibfnamefont {C.-S.}\ \bibnamefont {Niu}},\ }\bibfield  {title} {\bibinfo {title} {Semiclassical fourier transform for quantum computation},\ }\href@noop {} {\bibfield  {journal} {\bibinfo  {journal} {Physical Review Letters}\ }\textbf {\bibinfo {volume} {76}},\ \bibinfo {pages} {3228} (\bibinfo {year} {1996})}\BibitemShut {NoStop}%
\bibitem [{\citenamefont {Higgins}\ \emph {et~al.}(2007)\citenamefont {Higgins}, \citenamefont {Berry}, \citenamefont {Bartlett}, \citenamefont {Wiseman},\ and\ \citenamefont {Pryde}}]{higgins2007entanglement}%
  \BibitemOpen
  \bibfield  {author} {\bibinfo {author} {\bibfnamefont {B.~L.}\ \bibnamefont {Higgins}}, \bibinfo {author} {\bibfnamefont {D.~W.}\ \bibnamefont {Berry}}, \bibinfo {author} {\bibfnamefont {S.~D.}\ \bibnamefont {Bartlett}}, \bibinfo {author} {\bibfnamefont {H.~M.}\ \bibnamefont {Wiseman}},\ and\ \bibinfo {author} {\bibfnamefont {G.~J.}\ \bibnamefont {Pryde}},\ }\bibfield  {title} {\bibinfo {title} {Entanglement-free heisenberg-limited phase estimation},\ }\href@noop {} {\bibfield  {journal} {\bibinfo  {journal} {Nature}\ }\textbf {\bibinfo {volume} {450}},\ \bibinfo {pages} {393} (\bibinfo {year} {2007})}\BibitemShut {NoStop}%
\bibitem [{\citenamefont {Lin}\ and\ \citenamefont {Tong}(2022)}]{lin2022heisenberg}%
  \BibitemOpen
  \bibfield  {author} {\bibinfo {author} {\bibfnamefont {L.}~\bibnamefont {Lin}}\ and\ \bibinfo {author} {\bibfnamefont {Y.}~\bibnamefont {Tong}},\ }\bibfield  {title} {\bibinfo {title} {Heisenberg-limited ground-state energy estimation for early fault-tolerant quantum computers},\ }\href@noop {} {\bibfield  {journal} {\bibinfo  {journal} {PRX Quantum}\ }\textbf {\bibinfo {volume} {3}},\ \bibinfo {pages} {010318} (\bibinfo {year} {2022})}\BibitemShut {NoStop}%
\bibitem [{\citenamefont {Ding}\ and\ \citenamefont {Lin}(2023)}]{ding2023even}%
  \BibitemOpen
  \bibfield  {author} {\bibinfo {author} {\bibfnamefont {Z.}~\bibnamefont {Ding}}\ and\ \bibinfo {author} {\bibfnamefont {L.}~\bibnamefont {Lin}},\ }\bibfield  {title} {\bibinfo {title} {Even shorter quantum circuit for phase estimation on early fault-tolerant quantum computers with applications to ground-state energy estimation},\ }\href@noop {} {\bibfield  {journal} {\bibinfo  {journal} {PRX Quantum}\ }\textbf {\bibinfo {volume} {4}},\ \bibinfo {pages} {020331} (\bibinfo {year} {2023})}\BibitemShut {NoStop}%
\bibitem [{\citenamefont {Dutkiewicz}\ \emph {et~al.}(2024)\citenamefont {Dutkiewicz}, \citenamefont {Polla}, \citenamefont {Scheurer}, \citenamefont {Gogolin}, \citenamefont {Huggins},\ and\ \citenamefont {O'Brien}}]{dutkiewicz2024errormitigationcircuitdivision}%
  \BibitemOpen
  \bibfield  {author} {\bibinfo {author} {\bibfnamefont {A.}~\bibnamefont {Dutkiewicz}}, \bibinfo {author} {\bibfnamefont {S.}~\bibnamefont {Polla}}, \bibinfo {author} {\bibfnamefont {M.}~\bibnamefont {Scheurer}}, \bibinfo {author} {\bibfnamefont {C.}~\bibnamefont {Gogolin}}, \bibinfo {author} {\bibfnamefont {W.~J.}\ \bibnamefont {Huggins}},\ and\ \bibinfo {author} {\bibfnamefont {T.~E.}\ \bibnamefont {O'Brien}},\ }\href {https://arxiv.org/abs/2410.05369} {\bibinfo {title} {Error mitigation and circuit division for early fault-tolerant quantum phase estimation}} (\bibinfo {year} {2024}),\ \Eprint {https://arxiv.org/abs/2410.05369} {arXiv:2410.05369 [quant-ph]} \BibitemShut {NoStop}%
\bibitem [{\citenamefont {Bharti}\ \emph {et~al.}(2022)\citenamefont {Bharti}, \citenamefont {Cervera-Lierta}, \citenamefont {Kyaw}, \citenamefont {Haug}, \citenamefont {Alperin-Lea}, \citenamefont {Anand}, \citenamefont {Degroote}, \citenamefont {Heimonen}, \citenamefont {Kottmann}, \citenamefont {Menke} \emph {et~al.}}]{bharti2022noisy}%
  \BibitemOpen
  \bibfield  {author} {\bibinfo {author} {\bibfnamefont {K.}~\bibnamefont {Bharti}}, \bibinfo {author} {\bibfnamefont {A.}~\bibnamefont {Cervera-Lierta}}, \bibinfo {author} {\bibfnamefont {T.~H.}\ \bibnamefont {Kyaw}}, \bibinfo {author} {\bibfnamefont {T.}~\bibnamefont {Haug}}, \bibinfo {author} {\bibfnamefont {S.}~\bibnamefont {Alperin-Lea}}, \bibinfo {author} {\bibfnamefont {A.}~\bibnamefont {Anand}}, \bibinfo {author} {\bibfnamefont {M.}~\bibnamefont {Degroote}}, \bibinfo {author} {\bibfnamefont {H.}~\bibnamefont {Heimonen}}, \bibinfo {author} {\bibfnamefont {J.~S.}\ \bibnamefont {Kottmann}}, \bibinfo {author} {\bibfnamefont {T.}~\bibnamefont {Menke}}, \emph {et~al.},\ }\bibfield  {title} {\bibinfo {title} {Noisy intermediate-scale quantum algorithms},\ }\href@noop {} {\bibfield  {journal} {\bibinfo  {journal} {Reviews of Modern Physics}\ }\textbf {\bibinfo {volume} {94}},\ \bibinfo {pages} {015004} (\bibinfo {year} {2022})}\BibitemShut {NoStop}%
\bibitem [{\citenamefont {Tong}(2022)}]{tong2022designing}%
  \BibitemOpen
  \bibfield  {author} {\bibinfo {author} {\bibfnamefont {Y.}~\bibnamefont {Tong}},\ }\bibfield  {title} {\bibinfo {title} {Designing algorithms for estimating ground state properties on early fault-tolerant quantum computers},\ }\href@noop {} {\bibfield  {journal} {\bibinfo  {journal} {Quantum Views}\ }\textbf {\bibinfo {volume} {6}},\ \bibinfo {pages} {65} (\bibinfo {year} {2022})}\BibitemShut {NoStop}%
\bibitem [{\citenamefont {Parrish}\ and\ \citenamefont {McMahon}(2019)}]{parrish2019quantum}%
  \BibitemOpen
  \bibfield  {author} {\bibinfo {author} {\bibfnamefont {R.~M.}\ \bibnamefont {Parrish}}\ and\ \bibinfo {author} {\bibfnamefont {P.~L.}\ \bibnamefont {McMahon}},\ }\bibfield  {title} {\bibinfo {title} {Quantum filter diagonalization: Quantum eigendecomposition without full quantum phase estimation},\ }\href@noop {} {\bibfield  {journal} {\bibinfo  {journal} {arXiv preprint arXiv:1909.08925}\ } (\bibinfo {year} {2019})}\BibitemShut {NoStop}%
\bibitem [{\citenamefont {Seki}\ and\ \citenamefont {Yunoki}(2021)}]{seki2021quantum}%
  \BibitemOpen
  \bibfield  {author} {\bibinfo {author} {\bibfnamefont {K.}~\bibnamefont {Seki}}\ and\ \bibinfo {author} {\bibfnamefont {S.}~\bibnamefont {Yunoki}},\ }\bibfield  {title} {\bibinfo {title} {Quantum power method by a superposition of time-evolved states},\ }\href@noop {} {\bibfield  {journal} {\bibinfo  {journal} {PRX Quantum}\ }\textbf {\bibinfo {volume} {2}},\ \bibinfo {pages} {010333} (\bibinfo {year} {2021})}\BibitemShut {NoStop}%
\bibitem [{\citenamefont {Motta}\ \emph {et~al.}(2019)\citenamefont {Motta}, \citenamefont {Sun}, \citenamefont {Tan}, \citenamefont {O’Rourke}, \citenamefont {Ye}, \citenamefont {Minnich}, \citenamefont {Brandão},\ and\ \citenamefont {Chan}}]{motta2020determining}%
  \BibitemOpen
  \bibfield  {author} {\bibinfo {author} {\bibfnamefont {M.}~\bibnamefont {Motta}}, \bibinfo {author} {\bibfnamefont {C.}~\bibnamefont {Sun}}, \bibinfo {author} {\bibfnamefont {A.~T.~K.}\ \bibnamefont {Tan}}, \bibinfo {author} {\bibfnamefont {M.~J.}\ \bibnamefont {O’Rourke}}, \bibinfo {author} {\bibfnamefont {E.}~\bibnamefont {Ye}}, \bibinfo {author} {\bibfnamefont {A.~J.}\ \bibnamefont {Minnich}}, \bibinfo {author} {\bibfnamefont {F.~G. S.~L.}\ \bibnamefont {Brandão}},\ and\ \bibinfo {author} {\bibfnamefont {G.~K.-L.}\ \bibnamefont {Chan}},\ }\bibfield  {title} {\bibinfo {title} {Determining eigenstates and thermal states on a quantum computer using quantum imaginary time evolution},\ }\href {https://doi.org/10.1038/s41567-019-0704-4} {\bibfield  {journal} {\bibinfo  {journal} {Nature Physics}\ }\textbf {\bibinfo {volume} {16}},\ \bibinfo {pages} {205–210} (\bibinfo {year} {2019})}\BibitemShut {NoStop}%
\bibitem [{\citenamefont {Peruzzo}\ \emph {et~al.}(2014)\citenamefont {Peruzzo}, \citenamefont {McClean}, \citenamefont {Shadbolt}, \citenamefont {Yung}, \citenamefont {Zhou}, \citenamefont {Love}, \citenamefont {Aspuru-Guzik},\ and\ \citenamefont {O’brien}}]{peruzzo2014variational}%
  \BibitemOpen
  \bibfield  {author} {\bibinfo {author} {\bibfnamefont {A.}~\bibnamefont {Peruzzo}}, \bibinfo {author} {\bibfnamefont {J.}~\bibnamefont {McClean}}, \bibinfo {author} {\bibfnamefont {P.}~\bibnamefont {Shadbolt}}, \bibinfo {author} {\bibfnamefont {M.-H.}\ \bibnamefont {Yung}}, \bibinfo {author} {\bibfnamefont {X.-Q.}\ \bibnamefont {Zhou}}, \bibinfo {author} {\bibfnamefont {P.~J.}\ \bibnamefont {Love}}, \bibinfo {author} {\bibfnamefont {A.}~\bibnamefont {Aspuru-Guzik}},\ and\ \bibinfo {author} {\bibfnamefont {J.~L.}\ \bibnamefont {O’brien}},\ }\bibfield  {title} {\bibinfo {title} {A variational eigenvalue solver on a photonic quantum processor},\ }\href {https://doi.org/10.1038/ncomms5213} {\bibfield  {journal} {\bibinfo  {journal} {Nature communications}\ }\textbf {\bibinfo {volume} {5}},\ \bibinfo {pages} {1} (\bibinfo {year} {2014})}\BibitemShut {NoStop}%
\bibitem [{\citenamefont {McClean}\ \emph {et~al.}(2018)\citenamefont {McClean}, \citenamefont {Boixo}, \citenamefont {Smelyanskiy}, \citenamefont {Babbush},\ and\ \citenamefont {Neven}}]{mcclean2018barren}%
  \BibitemOpen
  \bibfield  {author} {\bibinfo {author} {\bibfnamefont {J.~R.}\ \bibnamefont {McClean}}, \bibinfo {author} {\bibfnamefont {S.}~\bibnamefont {Boixo}}, \bibinfo {author} {\bibfnamefont {V.~N.}\ \bibnamefont {Smelyanskiy}}, \bibinfo {author} {\bibfnamefont {R.}~\bibnamefont {Babbush}},\ and\ \bibinfo {author} {\bibfnamefont {H.}~\bibnamefont {Neven}},\ }\bibfield  {title} {\bibinfo {title} {Barren plateaus in quantum neural network training landscapes},\ }\href@noop {} {\bibfield  {journal} {\bibinfo  {journal} {Nature communications}\ }\textbf {\bibinfo {volume} {9}},\ \bibinfo {pages} {4812} (\bibinfo {year} {2018})}\BibitemShut {NoStop}%
\bibitem [{\citenamefont {Saad}(1980)}]{saad1980rates}%
  \BibitemOpen
  \bibfield  {author} {\bibinfo {author} {\bibfnamefont {Y.}~\bibnamefont {Saad}},\ }\bibfield  {title} {\bibinfo {title} {On the rates of convergence of the lanczos and the block-lanczos methods},\ }\href@noop {} {\bibfield  {journal} {\bibinfo  {journal} {SIAM Journal on Numerical Analysis}\ }\textbf {\bibinfo {volume} {17}},\ \bibinfo {pages} {687} (\bibinfo {year} {1980})}\BibitemShut {NoStop}%
\bibitem [{\citenamefont {Epperly}\ \emph {et~al.}(2022)\citenamefont {Epperly}, \citenamefont {Lin},\ and\ \citenamefont {Nakatsukasa}}]{epperly2022theory}%
  \BibitemOpen
  \bibfield  {author} {\bibinfo {author} {\bibfnamefont {E.~N.}\ \bibnamefont {Epperly}}, \bibinfo {author} {\bibfnamefont {L.}~\bibnamefont {Lin}},\ and\ \bibinfo {author} {\bibfnamefont {Y.}~\bibnamefont {Nakatsukasa}},\ }\bibfield  {title} {\bibinfo {title} {A theory of quantum subspace diagonalization},\ }\href@noop {} {\bibfield  {journal} {\bibinfo  {journal} {SIAM Journal on Matrix Analysis and Applications}\ }\textbf {\bibinfo {volume} {43}},\ \bibinfo {pages} {1263} (\bibinfo {year} {2022})}\BibitemShut {NoStop}%
\bibitem [{\citenamefont {Yoshioka}\ \emph {et~al.}(2024)\citenamefont {Yoshioka}, \citenamefont {Amico}, \citenamefont {Kirby}, \citenamefont {Jurcevic}, \citenamefont {Dutt}, \citenamefont {Fuller}, \citenamefont {Garion}, \citenamefont {Haas}, \citenamefont {Hamamura}, \citenamefont {Ivrii} \emph {et~al.}}]{yoshioka2024diagonalization}%
  \BibitemOpen
  \bibfield  {author} {\bibinfo {author} {\bibfnamefont {N.}~\bibnamefont {Yoshioka}}, \bibinfo {author} {\bibfnamefont {M.}~\bibnamefont {Amico}}, \bibinfo {author} {\bibfnamefont {W.}~\bibnamefont {Kirby}}, \bibinfo {author} {\bibfnamefont {P.}~\bibnamefont {Jurcevic}}, \bibinfo {author} {\bibfnamefont {A.}~\bibnamefont {Dutt}}, \bibinfo {author} {\bibfnamefont {B.}~\bibnamefont {Fuller}}, \bibinfo {author} {\bibfnamefont {S.}~\bibnamefont {Garion}}, \bibinfo {author} {\bibfnamefont {H.}~\bibnamefont {Haas}}, \bibinfo {author} {\bibfnamefont {I.}~\bibnamefont {Hamamura}}, \bibinfo {author} {\bibfnamefont {A.}~\bibnamefont {Ivrii}}, \emph {et~al.},\ }\bibfield  {title} {\bibinfo {title} {Diagonalization of large many-body hamiltonians on a quantum processor},\ }\href@noop {} {\bibfield  {journal} {\bibinfo  {journal} {arXiv preprint arXiv:2407.14431}\ } (\bibinfo {year} {2024})}\BibitemShut {NoStop}%
\bibitem [{\citenamefont {Kirby}(2024)}]{Kirby_2024}%
  \BibitemOpen
  \bibfield  {author} {\bibinfo {author} {\bibfnamefont {W.}~\bibnamefont {Kirby}},\ }\bibfield  {title} {\bibinfo {title} {Analysis of quantum krylov algorithms with errors},\ }\href {https://doi.org/10.22331/q-2024-08-29-1457} {\bibfield  {journal} {\bibinfo  {journal} {Quantum}\ }\textbf {\bibinfo {volume} {8}},\ \bibinfo {pages} {1457} (\bibinfo {year} {2024})}\BibitemShut {NoStop}%
\bibitem [{\citenamefont {Lee}\ \emph {et~al.}(2024)\citenamefont {Lee}, \citenamefont {Lee},\ and\ \citenamefont {Huh}}]{Lee_2024}%
  \BibitemOpen
  \bibfield  {author} {\bibinfo {author} {\bibfnamefont {G.}~\bibnamefont {Lee}}, \bibinfo {author} {\bibfnamefont {D.}~\bibnamefont {Lee}},\ and\ \bibinfo {author} {\bibfnamefont {J.}~\bibnamefont {Huh}},\ }\bibfield  {title} {\bibinfo {title} {Sampling error analysis in quantum krylov subspace diagonalization},\ }\href {https://doi.org/10.22331/q-2024-09-19-1477} {\bibfield  {journal} {\bibinfo  {journal} {Quantum}\ }\textbf {\bibinfo {volume} {8}},\ \bibinfo {pages} {1477} (\bibinfo {year} {2024})}\BibitemShut {NoStop}%
\bibitem [{\citenamefont {Brato{\v{z}}}(1958)}]{bratovz1958calcul}%
  \BibitemOpen
  \bibfield  {author} {\bibinfo {author} {\bibfnamefont {S.}~\bibnamefont {Brato{\v{z}}}},\ }\bibfield  {title} {\bibinfo {title} {Le calcul non empirique des constantes de force et des d{\'e}riv{\'e}es du moment dipolaire},\ }in\ \href@noop {} {\emph {\bibinfo {booktitle} {Colloq. Int. CNRS}}},\ Vol.~\bibinfo {volume} {82}\ (\bibinfo {year} {1958})\ pp.\ \bibinfo {pages} {287--301}\BibitemShut {NoStop}%
\bibitem [{\citenamefont {Gerratt}\ and\ \citenamefont {Mills}(1968)}]{gerratt1968force}%
  \BibitemOpen
  \bibfield  {author} {\bibinfo {author} {\bibfnamefont {J.}~\bibnamefont {Gerratt}}\ and\ \bibinfo {author} {\bibfnamefont {I.~M.}\ \bibnamefont {Mills}},\ }\bibfield  {title} {\bibinfo {title} {Force constants and dipole-moment derivatives of molecules from perturbed hartree--fock calculations. i},\ }\href@noop {} {\bibfield  {journal} {\bibinfo  {journal} {The Journal of Chemical Physics}\ }\textbf {\bibinfo {volume} {49}},\ \bibinfo {pages} {1719} (\bibinfo {year} {1968})}\BibitemShut {NoStop}%
\bibitem [{\citenamefont {Pulay}(1969)}]{pulay1969ab}%
  \BibitemOpen
  \bibfield  {author} {\bibinfo {author} {\bibfnamefont {P.}~\bibnamefont {Pulay}},\ }\bibfield  {title} {\bibinfo {title} {Ab initio calculation of force constants and equilibrium geometries in polyatomic molecules: I. theory},\ }\href@noop {} {\bibfield  {journal} {\bibinfo  {journal} {Molecular Physics}\ }\textbf {\bibinfo {volume} {17}},\ \bibinfo {pages} {197} (\bibinfo {year} {1969})}\BibitemShut {NoStop}%
\bibitem [{\citenamefont {Kato}\ and\ \citenamefont {Morokuma}(1979)}]{kato1979energy}%
  \BibitemOpen
  \bibfield  {author} {\bibinfo {author} {\bibfnamefont {S.}~\bibnamefont {Kato}}\ and\ \bibinfo {author} {\bibfnamefont {K.}~\bibnamefont {Morokuma}},\ }\bibfield  {title} {\bibinfo {title} {Energy gradient in a multi-configurational scf formalism and its application to geometry optimization of trimethylene diradicals},\ }\href@noop {} {\bibfield  {journal} {\bibinfo  {journal} {Chemical Physics Letters}\ }\textbf {\bibinfo {volume} {65}},\ \bibinfo {pages} {19} (\bibinfo {year} {1979})}\BibitemShut {NoStop}%
\bibitem [{\citenamefont {Goddard}\ \emph {et~al.}(1979)\citenamefont {Goddard}, \citenamefont {Handy},\ and\ \citenamefont {Schaefer}}]{goddard1979gradient}%
  \BibitemOpen
  \bibfield  {author} {\bibinfo {author} {\bibfnamefont {J.~D.}\ \bibnamefont {Goddard}}, \bibinfo {author} {\bibfnamefont {N.~C.}\ \bibnamefont {Handy}},\ and\ \bibinfo {author} {\bibfnamefont {H.~F.}\ \bibnamefont {Schaefer}},\ }\bibfield  {title} {\bibinfo {title} {Gradient techniques for open-shell restricted hartree--fock and multiconfiguration self-consistent-field methods},\ }\href@noop {} {\bibfield  {journal} {\bibinfo  {journal} {The Journal of Chemical Physics}\ }\textbf {\bibinfo {volume} {71}},\ \bibinfo {pages} {1525} (\bibinfo {year} {1979})}\BibitemShut {NoStop}%
\bibitem [{\citenamefont {Pulay}\ \emph {et~al.}(1979)\citenamefont {Pulay}, \citenamefont {Fogarasi}, \citenamefont {Pang},\ and\ \citenamefont {Boggs}}]{pulay1979systematic}%
  \BibitemOpen
  \bibfield  {author} {\bibinfo {author} {\bibfnamefont {P.}~\bibnamefont {Pulay}}, \bibinfo {author} {\bibfnamefont {G.}~\bibnamefont {Fogarasi}}, \bibinfo {author} {\bibfnamefont {F.}~\bibnamefont {Pang}},\ and\ \bibinfo {author} {\bibfnamefont {J.~E.}\ \bibnamefont {Boggs}},\ }\bibfield  {title} {\bibinfo {title} {Systematic ab initio gradient calculation of molecular geometries, force constants, and dipole moment derivatives},\ }\href@noop {} {\bibfield  {journal} {\bibinfo  {journal} {Journal of the American Chemical Society}\ }\textbf {\bibinfo {volume} {101}},\ \bibinfo {pages} {2550} (\bibinfo {year} {1979})}\BibitemShut {NoStop}%
\bibitem [{\citenamefont {Brooks}\ \emph {et~al.}(1980)\citenamefont {Brooks}, \citenamefont {Laidig}, \citenamefont {Saxe}, \citenamefont {Goddard}, \citenamefont {Yamaguchi},\ and\ \citenamefont {Schaefer}}]{brooks1980analytic}%
  \BibitemOpen
  \bibfield  {author} {\bibinfo {author} {\bibfnamefont {B.~R.}\ \bibnamefont {Brooks}}, \bibinfo {author} {\bibfnamefont {W.~D.}\ \bibnamefont {Laidig}}, \bibinfo {author} {\bibfnamefont {P.}~\bibnamefont {Saxe}}, \bibinfo {author} {\bibfnamefont {J.~D.}\ \bibnamefont {Goddard}}, \bibinfo {author} {\bibfnamefont {Y.}~\bibnamefont {Yamaguchi}},\ and\ \bibinfo {author} {\bibfnamefont {H.~F.}\ \bibnamefont {Schaefer}},\ }\bibfield  {title} {\bibinfo {title} {Analytic gradients from correlated wave functions via the two-particle density matrix and the unitary group approach},\ }\href@noop {} {\bibfield  {journal} {\bibinfo  {journal} {The Journal of Chemical Physics}\ }\textbf {\bibinfo {volume} {72}},\ \bibinfo {pages} {4652} (\bibinfo {year} {1980})}\BibitemShut {NoStop}%
\bibitem [{\citenamefont {Krishnan}\ \emph {et~al.}(1980)\citenamefont {Krishnan}, \citenamefont {Schlegel},\ and\ \citenamefont {Pople}}]{krishnan1980derivative}%
  \BibitemOpen
  \bibfield  {author} {\bibinfo {author} {\bibfnamefont {R.}~\bibnamefont {Krishnan}}, \bibinfo {author} {\bibfnamefont {H.}~\bibnamefont {Schlegel}},\ and\ \bibinfo {author} {\bibfnamefont {J.}~\bibnamefont {Pople}},\ }\bibfield  {title} {\bibinfo {title} {Derivative studies in configuration--interaction theory},\ }\href@noop {} {\bibfield  {journal} {\bibinfo  {journal} {The Journal of Chemical Physics}\ }\textbf {\bibinfo {volume} {72}},\ \bibinfo {pages} {4654} (\bibinfo {year} {1980})}\BibitemShut {NoStop}%
\bibitem [{\citenamefont {Dupuis}(1981)}]{dupuis1981energy}%
  \BibitemOpen
  \bibfield  {author} {\bibinfo {author} {\bibfnamefont {M.}~\bibnamefont {Dupuis}},\ }\bibfield  {title} {\bibinfo {title} {Energy derivatives for configuration interaction wave functions},\ }\href@noop {} {\bibfield  {journal} {\bibinfo  {journal} {The Journal of Chemical Physics}\ }\textbf {\bibinfo {volume} {74}},\ \bibinfo {pages} {5758} (\bibinfo {year} {1981})}\BibitemShut {NoStop}%
\bibitem [{\citenamefont {Nakatsuji}\ \emph {et~al.}(1981)\citenamefont {Nakatsuji}, \citenamefont {Hayakawa},\ and\ \citenamefont {Hada}}]{nakatsuji1981force}%
  \BibitemOpen
  \bibfield  {author} {\bibinfo {author} {\bibfnamefont {H.}~\bibnamefont {Nakatsuji}}, \bibinfo {author} {\bibfnamefont {T.}~\bibnamefont {Hayakawa}},\ and\ \bibinfo {author} {\bibfnamefont {M.}~\bibnamefont {Hada}},\ }\bibfield  {title} {\bibinfo {title} {Force in scf theories. mc scf and open-shell rhf theories},\ }\href@noop {} {\bibfield  {journal} {\bibinfo  {journal} {Chemical Physics Letters}\ }\textbf {\bibinfo {volume} {80}},\ \bibinfo {pages} {94} (\bibinfo {year} {1981})}\BibitemShut {NoStop}%
\bibitem [{\citenamefont {Schaefer~III}\ and\ \citenamefont {Yamaguchi}(1986)}]{schaefer1986robert}%
  \BibitemOpen
  \bibfield  {author} {\bibinfo {author} {\bibfnamefont {H.}~\bibnamefont {Schaefer~III}}\ and\ \bibinfo {author} {\bibfnamefont {Y.}~\bibnamefont {Yamaguchi}},\ }\bibfield  {title} {\bibinfo {title} {Robert s. mulliken issue},\ }\href@noop {} {\bibfield  {journal} {\bibinfo  {journal} {J. Mol. Struct}\ }\textbf {\bibinfo {volume} {135}},\ \bibinfo {pages} {369} (\bibinfo {year} {1986})}\BibitemShut {NoStop}%
\bibitem [{\citenamefont {Kassal}\ and\ \citenamefont {Aspuru-Guzik}(2009)}]{kassal2009quantum}%
  \BibitemOpen
  \bibfield  {author} {\bibinfo {author} {\bibfnamefont {I.}~\bibnamefont {Kassal}}\ and\ \bibinfo {author} {\bibfnamefont {A.}~\bibnamefont {Aspuru-Guzik}},\ }\bibfield  {title} {\bibinfo {title} {Quantum algorithm for molecular properties and geometry optimization},\ }\href@noop {} {\bibfield  {journal} {\bibinfo  {journal} {The Journal of chemical physics}\ }\textbf {\bibinfo {volume} {131}} (\bibinfo {year} {2009})}\BibitemShut {NoStop}%
\bibitem [{\citenamefont {Parrish}\ \emph {et~al.}(2019)\citenamefont {Parrish}, \citenamefont {Hohenstein}, \citenamefont {McMahon},\ and\ \citenamefont {Martinez}}]{parrish2019hybrid}%
  \BibitemOpen
  \bibfield  {author} {\bibinfo {author} {\bibfnamefont {R.~M.}\ \bibnamefont {Parrish}}, \bibinfo {author} {\bibfnamefont {E.~G.}\ \bibnamefont {Hohenstein}}, \bibinfo {author} {\bibfnamefont {P.~L.}\ \bibnamefont {McMahon}},\ and\ \bibinfo {author} {\bibfnamefont {T.~J.}\ \bibnamefont {Martinez}},\ }\bibfield  {title} {\bibinfo {title} {Hybrid quantum/classical derivative theory: Analytical gradients and excited-state dynamics for the multistate contracted variational quantum eigensolver},\ }\href@noop {} {\bibfield  {journal} {\bibinfo  {journal} {arXiv preprint arXiv:1906.08728}\ } (\bibinfo {year} {2019})}\BibitemShut {NoStop}%
\bibitem [{\citenamefont {Hohenstein}\ \emph {et~al.}(2023)\citenamefont {Hohenstein}, \citenamefont {Oumarou}, \citenamefont {Al-Saadon}, \citenamefont {Anselmetti}, \citenamefont {Scheurer}, \citenamefont {Gogolin},\ and\ \citenamefont {Parrish}}]{hohenstein2022efficient}%
  \BibitemOpen
  \bibfield  {author} {\bibinfo {author} {\bibfnamefont {E.~G.}\ \bibnamefont {Hohenstein}}, \bibinfo {author} {\bibfnamefont {O.}~\bibnamefont {Oumarou}}, \bibinfo {author} {\bibfnamefont {R.}~\bibnamefont {Al-Saadon}}, \bibinfo {author} {\bibfnamefont {G.-L.~R.}\ \bibnamefont {Anselmetti}}, \bibinfo {author} {\bibfnamefont {M.}~\bibnamefont {Scheurer}}, \bibinfo {author} {\bibfnamefont {C.}~\bibnamefont {Gogolin}},\ and\ \bibinfo {author} {\bibfnamefont {R.~M.}\ \bibnamefont {Parrish}},\ }\bibfield  {title} {\bibinfo {title} {{Efficient quantum analytic nuclear gradients with double factorization}},\ }\href {https://doi.org/10.1063/5.0137167} {\bibfield  {journal} {\bibinfo  {journal} {The Journal of Chemical Physics}\ }\textbf {\bibinfo {volume} {158}},\ \bibinfo {pages} {114119} (\bibinfo {year} {2023})}\BibitemShut {NoStop}%
\bibitem [{\citenamefont {O’Brien}\ \emph {et~al.}(2019)\citenamefont {O’Brien}, \citenamefont {Senjean}, \citenamefont {Sagastizabal}, \citenamefont {Bonet-Monroig}, \citenamefont {Dutkiewicz}, \citenamefont {Buda}, \citenamefont {DiCarlo},\ and\ \citenamefont {Visscher}}]{o2019calculating}%
  \BibitemOpen
  \bibfield  {author} {\bibinfo {author} {\bibfnamefont {T.~E.}\ \bibnamefont {O’Brien}}, \bibinfo {author} {\bibfnamefont {B.}~\bibnamefont {Senjean}}, \bibinfo {author} {\bibfnamefont {R.}~\bibnamefont {Sagastizabal}}, \bibinfo {author} {\bibfnamefont {X.}~\bibnamefont {Bonet-Monroig}}, \bibinfo {author} {\bibfnamefont {A.}~\bibnamefont {Dutkiewicz}}, \bibinfo {author} {\bibfnamefont {F.}~\bibnamefont {Buda}}, \bibinfo {author} {\bibfnamefont {L.}~\bibnamefont {DiCarlo}},\ and\ \bibinfo {author} {\bibfnamefont {L.}~\bibnamefont {Visscher}},\ }\bibfield  {title} {\bibinfo {title} {Calculating energy derivatives for quantum chemistry on a quantum computer},\ }\bibfield  {journal} {\bibinfo  {journal} {npj Quantum Information}\ }\textbf {\bibinfo {volume} {5}},\ \href {https://doi.org/10.1038/s41534-019-0213-4} {10.1038/s41534-019-0213-4} (\bibinfo {year} {2019})\BibitemShut {NoStop}%
\bibitem [{\citenamefont {O’Brien}\ \emph {et~al.}(2022)\citenamefont {O’Brien}, \citenamefont {Streif}, \citenamefont {Rubin}, \citenamefont {Santagati}, \citenamefont {Su}, \citenamefont {Huggins}, \citenamefont {Goings}, \citenamefont {Moll}, \citenamefont {Kyoseva}, \citenamefont {Degroote}, \citenamefont {Tautermann}, \citenamefont {Lee}, \citenamefont {Berry}, \citenamefont {Wiebe},\ and\ \citenamefont {Babbush}}]{o2022efficient}%
  \BibitemOpen
  \bibfield  {author} {\bibinfo {author} {\bibfnamefont {T.~E.}\ \bibnamefont {O’Brien}}, \bibinfo {author} {\bibfnamefont {M.}~\bibnamefont {Streif}}, \bibinfo {author} {\bibfnamefont {N.~C.}\ \bibnamefont {Rubin}}, \bibinfo {author} {\bibfnamefont {R.}~\bibnamefont {Santagati}}, \bibinfo {author} {\bibfnamefont {Y.}~\bibnamefont {Su}}, \bibinfo {author} {\bibfnamefont {W.~J.}\ \bibnamefont {Huggins}}, \bibinfo {author} {\bibfnamefont {J.~J.}\ \bibnamefont {Goings}}, \bibinfo {author} {\bibfnamefont {N.}~\bibnamefont {Moll}}, \bibinfo {author} {\bibfnamefont {E.}~\bibnamefont {Kyoseva}}, \bibinfo {author} {\bibfnamefont {M.}~\bibnamefont {Degroote}}, \bibinfo {author} {\bibfnamefont {C.~S.}\ \bibnamefont {Tautermann}}, \bibinfo {author} {\bibfnamefont {J.}~\bibnamefont {Lee}}, \bibinfo {author} {\bibfnamefont {D.~W.}\ \bibnamefont {Berry}}, \bibinfo {author} {\bibfnamefont {N.}~\bibnamefont {Wiebe}},\ and\ \bibinfo {author} {\bibfnamefont {R.}~\bibnamefont {Babbush}},\ }\bibfield  {title} {\bibinfo {title}
  {Efficient quantum computation of molecular forces and other energy gradients},\ }\bibfield  {journal} {\bibinfo  {journal} {Physical Review Research}\ }\textbf {\bibinfo {volume} {4}},\ \href {https://doi.org/10.1103/physrevresearch.4.043210} {10.1103/physrevresearch.4.043210} (\bibinfo {year} {2022})\BibitemShut {NoStop}%
\bibitem [{\citenamefont {Mitarai}\ \emph {et~al.}(2020)\citenamefont {Mitarai}, \citenamefont {Nakagawa},\ and\ \citenamefont {Mizukami}}]{mitarai2020theory}%
  \BibitemOpen
  \bibfield  {author} {\bibinfo {author} {\bibfnamefont {K.}~\bibnamefont {Mitarai}}, \bibinfo {author} {\bibfnamefont {Y.~O.}\ \bibnamefont {Nakagawa}},\ and\ \bibinfo {author} {\bibfnamefont {W.}~\bibnamefont {Mizukami}},\ }\bibfield  {title} {\bibinfo {title} {Theory of analytical energy derivatives for the variational quantum eigensolver},\ }\href@noop {} {\bibfield  {journal} {\bibinfo  {journal} {Physical Review Research}\ }\textbf {\bibinfo {volume} {2}},\ \bibinfo {pages} {013129} (\bibinfo {year} {2020})}\BibitemShut {NoStop}%
\bibitem [{\citenamefont {Parrish}\ \emph {et~al.}(2021)\citenamefont {Parrish}, \citenamefont {Anselmetti},\ and\ \citenamefont {Gogolin}}]{parrish2021analytical}%
  \BibitemOpen
  \bibfield  {author} {\bibinfo {author} {\bibfnamefont {R.~M.}\ \bibnamefont {Parrish}}, \bibinfo {author} {\bibfnamefont {G.-L.~R.}\ \bibnamefont {Anselmetti}},\ and\ \bibinfo {author} {\bibfnamefont {C.}~\bibnamefont {Gogolin}},\ }\bibfield  {title} {\bibinfo {title} {Analytical ground-and excited-state gradients for molecular electronic structure theory from hybrid quantum/classical methods},\ }\href@noop {} {\bibfield  {journal} {\bibinfo  {journal} {arXiv preprint arXiv:2110.05040}\ } (\bibinfo {year} {2021})}\BibitemShut {NoStop}%
\bibitem [{\citenamefont {Kirby}\ \emph {et~al.}(2023)\citenamefont {Kirby}, \citenamefont {Motta},\ and\ \citenamefont {Mezzacapo}}]{kirby2023exact}%
  \BibitemOpen
  \bibfield  {author} {\bibinfo {author} {\bibfnamefont {W.}~\bibnamefont {Kirby}}, \bibinfo {author} {\bibfnamefont {M.}~\bibnamefont {Motta}},\ and\ \bibinfo {author} {\bibfnamefont {A.}~\bibnamefont {Mezzacapo}},\ }\bibfield  {title} {\bibinfo {title} {Exact and efficient lanczos method on a quantum computer},\ }\href@noop {} {\bibfield  {journal} {\bibinfo  {journal} {Quantum}\ }\textbf {\bibinfo {volume} {7}},\ \bibinfo {pages} {1018} (\bibinfo {year} {2023})}\BibitemShut {NoStop}%
\bibitem [{\citenamefont {Zhang}\ \emph {et~al.}(2024)\citenamefont {Zhang}, \citenamefont {Wang}, \citenamefont {Xu},\ and\ \citenamefont {Li}}]{Zhang_2024}%
  \BibitemOpen
  \bibfield  {author} {\bibinfo {author} {\bibfnamefont {Z.}~\bibnamefont {Zhang}}, \bibinfo {author} {\bibfnamefont {A.}~\bibnamefont {Wang}}, \bibinfo {author} {\bibfnamefont {X.}~\bibnamefont {Xu}},\ and\ \bibinfo {author} {\bibfnamefont {Y.}~\bibnamefont {Li}},\ }\bibfield  {title} {\bibinfo {title} {Measurement-efficient quantum krylov subspace diagonalisation},\ }\href {https://doi.org/10.22331/q-2024-08-13-1438} {\bibfield  {journal} {\bibinfo  {journal} {Quantum}\ }\textbf {\bibinfo {volume} {8}},\ \bibinfo {pages} {1438} (\bibinfo {year} {2024})}\BibitemShut {NoStop}%
\bibitem [{\citenamefont {L{\"o}wdin}(1956)}]{lowdin1956quantum}%
  \BibitemOpen
  \bibfield  {author} {\bibinfo {author} {\bibfnamefont {P.-O.}\ \bibnamefont {L{\"o}wdin}},\ }\bibfield  {title} {\bibinfo {title} {Quantum theory of cohesive properties of solids},\ }\href@noop {} {\bibfield  {journal} {\bibinfo  {journal} {Advances in Physics}\ }\textbf {\bibinfo {volume} {5}},\ \bibinfo {pages} {1} (\bibinfo {year} {1956})}\BibitemShut {NoStop}%
\bibitem [{\citenamefont {L{\"o}wdin}(1970)}]{lowdin1970nonorthogonality}%
  \BibitemOpen
  \bibfield  {author} {\bibinfo {author} {\bibfnamefont {P.-O.}\ \bibnamefont {L{\"o}wdin}},\ }\bibfield  {title} {\bibinfo {title} {On the nonorthogonality problem},\ }in\ \href@noop {} {\emph {\bibinfo {booktitle} {Advances in quantum chemistry}}},\ Vol.~\bibinfo {volume} {5}\ (\bibinfo  {publisher} {Elsevier},\ \bibinfo {year} {1970})\ pp.\ \bibinfo {pages} {185--199}\BibitemShut {NoStop}%
\bibitem [{\citenamefont {Pople}\ \emph {et~al.}(1979)\citenamefont {Pople}, \citenamefont {Krishnan}, \citenamefont {Schlegel},\ and\ \citenamefont {Binkley}}]{hf_grad}%
  \BibitemOpen
  \bibfield  {author} {\bibinfo {author} {\bibfnamefont {J.~A.}\ \bibnamefont {Pople}}, \bibinfo {author} {\bibfnamefont {R.}~\bibnamefont {Krishnan}}, \bibinfo {author} {\bibfnamefont {H.~B.}\ \bibnamefont {Schlegel}},\ and\ \bibinfo {author} {\bibfnamefont {J.~S.}\ \bibnamefont {Binkley}},\ }\bibfield  {title} {\bibinfo {title} {Derivative studies in hartree-fock and møller-plesset theories},\ }\href {https://doi.org/https://doi.org/10.1002/qua.560160825} {\bibfield  {journal} {\bibinfo  {journal} {International Journal of Quantum Chemistry}\ }\textbf {\bibinfo {volume} {16}},\ \bibinfo {pages} {225} (\bibinfo {year} {1979})},\ \Eprint {https://arxiv.org/abs/https://onlinelibrary.wiley.com/doi/pdf/10.1002/qua.560160825} {https://onlinelibrary.wiley.com/doi/pdf/10.1002/qua.560160825} \BibitemShut {NoStop}%
\bibitem [{\citenamefont {Koch}(2011)}]{koch20118}%
  \BibitemOpen
  \bibfield  {author} {\bibinfo {author} {\bibfnamefont {E.}~\bibnamefont {Koch}},\ }\bibfield  {title} {\bibinfo {title} {8 the lanczos method},\ }\href@noop {} {\bibfield  {journal} {\bibinfo  {journal} {The LDA+ DMFT approach to strongly correlated materials}\ } (\bibinfo {year} {2011})}\BibitemShut {NoStop}%
\bibitem [{\citenamefont {Dong}\ \emph {et~al.}(2021)\citenamefont {Dong}, \citenamefont {Meng}, \citenamefont {Whaley},\ and\ \citenamefont {Lin}}]{dong2021efficient}%
  \BibitemOpen
  \bibfield  {author} {\bibinfo {author} {\bibfnamefont {Y.}~\bibnamefont {Dong}}, \bibinfo {author} {\bibfnamefont {X.}~\bibnamefont {Meng}}, \bibinfo {author} {\bibfnamefont {K.~B.}\ \bibnamefont {Whaley}},\ and\ \bibinfo {author} {\bibfnamefont {L.}~\bibnamefont {Lin}},\ }\bibfield  {title} {\bibinfo {title} {Efficient phase-factor evaluation in quantum signal processing},\ }\href@noop {} {\bibfield  {journal} {\bibinfo  {journal} {Physical Review A}\ }\textbf {\bibinfo {volume} {103}},\ \bibinfo {pages} {042419} (\bibinfo {year} {2021})}\BibitemShut {NoStop}%
\bibitem [{\citenamefont {Motlagh}\ and\ \citenamefont {Wiebe}(2024)}]{Motlagh_2024}%
  \BibitemOpen
  \bibfield  {author} {\bibinfo {author} {\bibfnamefont {D.}~\bibnamefont {Motlagh}}\ and\ \bibinfo {author} {\bibfnamefont {N.}~\bibnamefont {Wiebe}},\ }\bibfield  {title} {\bibinfo {title} {Generalized quantum signal processing},\ }\bibfield  {journal} {\bibinfo  {journal} {PRX Quantum}\ }\textbf {\bibinfo {volume} {5}},\ \href {https://doi.org/10.1103/prxquantum.5.020368} {10.1103/prxquantum.5.020368} (\bibinfo {year} {2024})\BibitemShut {NoStop}%
\bibitem [{\citenamefont {Gily{\'e}n}\ \emph {et~al.}(2019)\citenamefont {Gily{\'e}n}, \citenamefont {Su}, \citenamefont {Low},\ and\ \citenamefont {Wiebe}}]{gilyen2019quantum}%
  \BibitemOpen
  \bibfield  {author} {\bibinfo {author} {\bibfnamefont {A.}~\bibnamefont {Gily{\'e}n}}, \bibinfo {author} {\bibfnamefont {Y.}~\bibnamefont {Su}}, \bibinfo {author} {\bibfnamefont {G.~H.}\ \bibnamefont {Low}},\ and\ \bibinfo {author} {\bibfnamefont {N.}~\bibnamefont {Wiebe}},\ }\bibfield  {title} {\bibinfo {title} {Quantum singular value transformation and beyond: exponential improvements for quantum matrix arithmetics},\ }in\ \href@noop {} {\emph {\bibinfo {booktitle} {Proceedings of the 51st Annual ACM SIGACT Symposium on Theory of Computing}}}\ (\bibinfo {year} {2019})\ pp.\ \bibinfo {pages} {193--204}\BibitemShut {NoStop}%
\bibitem [{\citenamefont {Oumarou}\ \emph {et~al.}(2024)\citenamefont {Oumarou}, \citenamefont {Scheurer}, \citenamefont {Parrish}, \citenamefont {Hohenstein},\ and\ \citenamefont {Gogolin}}]{Oumarou_2024}%
  \BibitemOpen
  \bibfield  {author} {\bibinfo {author} {\bibfnamefont {O.}~\bibnamefont {Oumarou}}, \bibinfo {author} {\bibfnamefont {M.}~\bibnamefont {Scheurer}}, \bibinfo {author} {\bibfnamefont {R.~M.}\ \bibnamefont {Parrish}}, \bibinfo {author} {\bibfnamefont {E.~G.}\ \bibnamefont {Hohenstein}},\ and\ \bibinfo {author} {\bibfnamefont {C.}~\bibnamefont {Gogolin}},\ }\bibfield  {title} {\bibinfo {title} {Accelerating quantum computations of chemistry through regularized compressed double factorization},\ }\href {https://doi.org/10.22331/q-2024-06-13-1371} {\bibfield  {journal} {\bibinfo  {journal} {Quantum}\ }\textbf {\bibinfo {volume} {8}},\ \bibinfo {pages} {1371} (\bibinfo {year} {2024})}\BibitemShut {NoStop}%
\bibitem [{\citenamefont {Huang}\ \emph {et~al.}(2020)\citenamefont {Huang}, \citenamefont {Kueng},\ and\ \citenamefont {Preskill}}]{huang2020predicting}%
  \BibitemOpen
  \bibfield  {author} {\bibinfo {author} {\bibfnamefont {H.-Y.}\ \bibnamefont {Huang}}, \bibinfo {author} {\bibfnamefont {R.}~\bibnamefont {Kueng}},\ and\ \bibinfo {author} {\bibfnamefont {J.}~\bibnamefont {Preskill}},\ }\bibfield  {title} {\bibinfo {title} {Predicting many properties of a quantum system from very few measurements},\ }\href {https://doi.org/10.1038/s41567-020-0932-7} {\bibfield  {journal} {\bibinfo  {journal} {Nature Physics}\ }\textbf {\bibinfo {volume} {16}},\ \bibinfo {pages} {1050} (\bibinfo {year} {2020})}\BibitemShut {NoStop}%
\bibitem [{\citenamefont {Zhao}\ \emph {et~al.}(2021)\citenamefont {Zhao}, \citenamefont {Rubin},\ and\ \citenamefont {Miyake}}]{zhao_fermionic_2021}%
  \BibitemOpen
  \bibfield  {author} {\bibinfo {author} {\bibfnamefont {A.}~\bibnamefont {Zhao}}, \bibinfo {author} {\bibfnamefont {N.~C.}\ \bibnamefont {Rubin}},\ and\ \bibinfo {author} {\bibfnamefont {A.}~\bibnamefont {Miyake}},\ }\bibfield  {title} {\bibinfo {title} {Fermionic partial tomography via classical shadows},\ }\href@noop {} {\bibfield  {journal} {\bibinfo  {journal} {Phys. Rev. Lett.}\ }\textbf {\bibinfo {volume} {127}},\ \bibinfo {pages} {110504} (\bibinfo {year} {2021})}\BibitemShut {NoStop}%
\bibitem [{\citenamefont {Low}(2022)}]{low_classical_2022}%
  \BibitemOpen
  \bibfield  {author} {\bibinfo {author} {\bibfnamefont {G.~H.}\ \bibnamefont {Low}},\ }\bibfield  {title} {\bibinfo {title} {Classical shadows of fermions with particle number symmetry},\ }\href@noop {} {\bibfield  {journal} {\bibinfo  {journal} {arXiv preprint}\ ,\ \bibinfo {pages} {arXiv:2208.08964}} (\bibinfo {year} {2022})}\BibitemShut {NoStop}%
\bibitem [{\citenamefont {Wan}\ \emph {et~al.}(2023)\citenamefont {Wan}, \citenamefont {Huggins}, \citenamefont {Lee},\ and\ \citenamefont {Babbush}}]{wan2022matchgate}%
  \BibitemOpen
  \bibfield  {author} {\bibinfo {author} {\bibfnamefont {K.}~\bibnamefont {Wan}}, \bibinfo {author} {\bibfnamefont {W.~J.}\ \bibnamefont {Huggins}}, \bibinfo {author} {\bibfnamefont {J.}~\bibnamefont {Lee}},\ and\ \bibinfo {author} {\bibfnamefont {R.}~\bibnamefont {Babbush}},\ }\bibfield  {title} {\bibinfo {title} {Matchgate shadows for fermionic quantum simulation},\ }\href {https://doi.org/10.1007/s00220-023-04844-0} {\bibfield  {journal} {\bibinfo  {journal} {Communications in Mathematical Physics}\ }\textbf {\bibinfo {volume} {404}},\ \bibinfo {pages} {629–700} (\bibinfo {year} {2023})}\BibitemShut {NoStop}%
\bibitem [{\citenamefont {Huggins}\ \emph {et~al.}(2022)\citenamefont {Huggins}, \citenamefont {O’Gorman}, \citenamefont {Rubin}, \citenamefont {Reichman}, \citenamefont {Babbush},\ and\ \citenamefont {Lee}}]{huggins_unbiasing_2022}%
  \BibitemOpen
  \bibfield  {author} {\bibinfo {author} {\bibfnamefont {W.~J.}\ \bibnamefont {Huggins}}, \bibinfo {author} {\bibfnamefont {B.~A.}\ \bibnamefont {O’Gorman}}, \bibinfo {author} {\bibfnamefont {N.~C.}\ \bibnamefont {Rubin}}, \bibinfo {author} {\bibfnamefont {D.~R.}\ \bibnamefont {Reichman}}, \bibinfo {author} {\bibfnamefont {R.}~\bibnamefont {Babbush}},\ and\ \bibinfo {author} {\bibfnamefont {J.}~\bibnamefont {Lee}},\ }\bibfield  {title} {\bibinfo {title} {Unbiasing fermionic quantum {Monte} {Carlo} with a quantum computer},\ }\href {https://doi.org/10.1038/s41586-021-04351-z} {\bibfield  {journal} {\bibinfo  {journal} {Nature}\ }\textbf {\bibinfo {volume} {603}},\ \bibinfo {pages} {416} (\bibinfo {year} {2022})}\BibitemShut {NoStop}%
\bibitem [{\citenamefont {Scheurer}\ \emph {et~al.}(2024)\citenamefont {Scheurer}, \citenamefont {Anselmetti}, \citenamefont {Oumarou}, \citenamefont {Gogolin},\ and\ \citenamefont {Rubin}}]{Scheurer_2024}%
  \BibitemOpen
  \bibfield  {author} {\bibinfo {author} {\bibfnamefont {M.}~\bibnamefont {Scheurer}}, \bibinfo {author} {\bibfnamefont {G.-L.~R.}\ \bibnamefont {Anselmetti}}, \bibinfo {author} {\bibfnamefont {O.}~\bibnamefont {Oumarou}}, \bibinfo {author} {\bibfnamefont {C.}~\bibnamefont {Gogolin}},\ and\ \bibinfo {author} {\bibfnamefont {N.~C.}\ \bibnamefont {Rubin}},\ }\bibfield  {title} {\bibinfo {title} {Tailored and externally corrected coupled cluster with quantum inputs},\ }\href {https://doi.org/10.1021/acs.jctc.4c00037} {\bibfield  {journal} {\bibinfo  {journal} {Journal of Chemical Theory and Computation}\ }\textbf {\bibinfo {volume} {20}},\ \bibinfo {pages} {5068–5093} (\bibinfo {year} {2024})}\BibitemShut {NoStop}%
\bibitem [{\citenamefont {Sun}\ \emph {et~al.}(2018)\citenamefont {Sun}, \citenamefont {Berkelbach}, \citenamefont {Blunt}, \citenamefont {Booth}, \citenamefont {Guo}, \citenamefont {Li}, \citenamefont {Liu}, \citenamefont {McClain}, \citenamefont {Sayfutyarova}, \citenamefont {Sharma} \emph {et~al.}}]{sun2018pyscf}%
  \BibitemOpen
  \bibfield  {author} {\bibinfo {author} {\bibfnamefont {Q.}~\bibnamefont {Sun}}, \bibinfo {author} {\bibfnamefont {T.~C.}\ \bibnamefont {Berkelbach}}, \bibinfo {author} {\bibfnamefont {N.~S.}\ \bibnamefont {Blunt}}, \bibinfo {author} {\bibfnamefont {G.~H.}\ \bibnamefont {Booth}}, \bibinfo {author} {\bibfnamefont {S.}~\bibnamefont {Guo}}, \bibinfo {author} {\bibfnamefont {Z.}~\bibnamefont {Li}}, \bibinfo {author} {\bibfnamefont {J.}~\bibnamefont {Liu}}, \bibinfo {author} {\bibfnamefont {J.~D.}\ \bibnamefont {McClain}}, \bibinfo {author} {\bibfnamefont {E.~R.}\ \bibnamefont {Sayfutyarova}}, \bibinfo {author} {\bibfnamefont {S.}~\bibnamefont {Sharma}}, \emph {et~al.},\ }\bibfield  {title} {\bibinfo {title} {Pyscf: the python-based simulations of chemistry framework},\ }\href {https://doi.org/10.1002/wcms.1340} {\bibfield  {journal} {\bibinfo  {journal} {Wiley Interdisciplinary Reviews: Computational Molecular Science}\ }\textbf {\bibinfo {volume} {8}},\ \bibinfo {pages} {e1340} (\bibinfo {year}
  {2018})}\BibitemShut {NoStop}%
\bibitem [{\citenamefont {Bergholm}\ \emph {et~al.}(2018)\citenamefont {Bergholm}, \citenamefont {Izaac}, \citenamefont {Schuld}, \citenamefont {Gogolin}, \citenamefont {Ahmed}, \citenamefont {Ajith}, \citenamefont {Alam}, \citenamefont {Alonso-Linaje}, \citenamefont {AkashNarayanan}, \citenamefont {Asadi} \emph {et~al.}}]{bergholm2018pennylane}%
  \BibitemOpen
  \bibfield  {author} {\bibinfo {author} {\bibfnamefont {V.}~\bibnamefont {Bergholm}}, \bibinfo {author} {\bibfnamefont {J.}~\bibnamefont {Izaac}}, \bibinfo {author} {\bibfnamefont {M.}~\bibnamefont {Schuld}}, \bibinfo {author} {\bibfnamefont {C.}~\bibnamefont {Gogolin}}, \bibinfo {author} {\bibfnamefont {S.}~\bibnamefont {Ahmed}}, \bibinfo {author} {\bibfnamefont {V.}~\bibnamefont {Ajith}}, \bibinfo {author} {\bibfnamefont {M.~S.}\ \bibnamefont {Alam}}, \bibinfo {author} {\bibfnamefont {G.}~\bibnamefont {Alonso-Linaje}}, \bibinfo {author} {\bibfnamefont {B.}~\bibnamefont {AkashNarayanan}}, \bibinfo {author} {\bibfnamefont {A.}~\bibnamefont {Asadi}}, \emph {et~al.},\ }\bibfield  {title} {\bibinfo {title} {Pennylane: Automatic differentiation of hybrid quantum-classical computations},\ }\href@noop {} {\bibfield  {journal} {\bibinfo  {journal} {arXiv preprint arXiv:1811.04968}\ } (\bibinfo {year} {2018})}\BibitemShut {NoStop}%
\end{thebibliography}%
